\documentclass[referee]{aa}
\usepackage{graphicx}
\usepackage{natbib}
\usepackage{wrapfig}
\usepackage{sidecap}
\usepackage{caption}
\usepackage{subcaption}
\usepackage{longtable}
\usepackage{rotating}
\usepackage{lscape}
\usepackage{multicol}
\usepackage{txfonts}

\newcommand\msun{\mathrm{M}_\odot}
   \title{Near-Infrared Time-Series Photometry in the Field of Cygnus OB2 Association
           }
   \subtitle{I - Rotational Scenario For Candidate Members}

   \author{J. Roquette
          \inst{1}
          \and
          J. Bouvier \inst{2}
          \and
          S.H.P. Alencar\inst{1}
          \and
          L.P.R. Vaz\inst{1}
          \and
          M. G. Guarcello\inst{3}
          }

   \institute{Departamento de F\'isica - ICEx - UFMG, Av. Ant\^onio Carlos, 6627, 30270-901 Belo Horizonte, MG, Brazil\\
              \email{roquette@fisica.ufmg.br}
         \and
             Univ. Grenoble Alpes, CNRS, IPAG, F-38000 Grenoble, France \\
             \and
              INAF - Osservatorio Astronomico di Palermo, Piazza del Parlamento 1, I-90134, Palermo, Italy\\
             }

   \date{Recieved December 23, 2016}
   \abstract 
     {In the last decades, the early pre main sequence stellar rotational evolution picture has been
       constrained by studies targetting different regions at a variety of ages with respect to
       young star formation.  Observational studies suggest a dependence of rotation with mass, and
       for some mass ranges a connection between rotation and the presence of a circumstellar disk.
     Not still fully explored, though, is the role of environmental conditions on the rotational
   regulation.}
   {We investigate the rotational properties of candidate members of the young massive association
   Cygnus OB2.  By evaluating their rotational properties, we address questions regarding the effect
 of environment properties on PMS rotational evolution. } 
   {We studied JHK-band variability in 5083 candidate members (24$\%$ of them are disk-bearing
     stars). The selection of variable stars was done using Stetson variability index, and period
     search was performed using Lomb-Scargle periodogram for periods between 0.83-45 days. Period
     detections were verified by using False Alarme Probability levels, the Saunders statistics,
     string/rope length method, and visual verification of folded light curves. 
   }
   {We identified 1224 periodic variable stars (24$\%$ of the candidate member sample, 8$\%$ of the
     disk-bearing sample, and 28$\%$ of the non-disked sample). Monte Carlo simulations were
     performed in order to evaluate completeness and contamination of the periodic sample, out of
     which 894 measured periods were considered reliable. Our study was considered reasonably
   complete for periods between 2 and 30 days.}
     {The general scenario for the rotational evolution of young stars seen in other regions is
       confirmed by Cygnus OB2 period distributions, with disked stars rotating on average slower
       than non-disked stars. A mass-rotation dependence was also verified, but as in NGC 6530, very
       low mass stars ($M\leq0.4\msun$) are rotating on average slower than higher mass stars
       ($0.4\;\msun<M\leq1.4\msun$). We observed an excess of slow rotators among the lower mass
       population. The disk and mass-rotation connection was also analysed by taking into account
       the incident UV radiation arising from O stars in the association. Results compatible with
       the disk-locking scenario were verified for stars with low UV incidence, but no statistical
       significant relation between rotation and disk presence was verified for stars with high UV
       incidence suggesting that massive stars can have an important role on regulating the rotation
       of nearby low mass stars.
     }

\begin{document} 
   \maketitle
\section{Introduction}
 
The angular momentum (AM) evolution during the early stages of the stellar life is one of the most
fundamental questions currently under debate \citep[e.g.,][]{Bodenheimer}. If AM conservation was
the sole responsible for the early rotational evolution during star contraction towards the zero age
main sequence (ZAMS) phase, pre main sequence (PMS) stars should reach their first few Myr with
spin rates close to critical values (when the centrifugal forces balance gravity). On the contrary,
observations mapping rotational velocities of low mass PMS stars \citep[hereafter T Tauri stars;
TTS,][]{1945Joy} found that their typical rotational velocities are only a fraction of their
critical velocities \citep[e.g.][]{1982Vogel,1986jbouvier}. This suggests that there must exist some
ongoing physical mechanisms on such stars that counteract the spin-up expected from stellar
contraction. 

The AM of a forming star is determined by both internal and external physical processes. Internal
processes determine how AM is transported in the stellar interior. The most popular scenario assumes
that a radiative core and a convective envelope rotate as solid bodies but with different angular
velocities \citep[e.g.][]{2013Gallet}. External processes are responsible for the AM loss
from the stellar surface, and models for such processes include the magnetic star-disk interaction
\citep[e.g.,][]{GhoshLamb1979,2005Matta}, accretion-powered stellar winds
\citep[e.g.,][]{Matt2012}, and mass ejections \citep[e.g., Conical
Winds:][]{2013Zanni,2009Romanova}. To constrain the physical models, recent studies have
been looking for correlations between stellar rotation rates and parameters like stellar mass,
circumstellar disk indicators, X-ray emission, and mass accretion rates. 

A controversial issue still in debate is the ``disk-locking'' process
\citep[e.g.,][]{GhoshLamb1979,Koenigl1991}, which is based on the observational evidence that
accreting stars are on average slower rotators than non-accreting stars. According to the
disk-locking scenario, PMS stars still magnetically interacting with their disks would be
prevented from spinning-up via star-disk interaction, despite the fact that they are contracting
towards the ZAMS.  Consequently, they would remain with almost constant rotational velocity during
their first few Myr. The disk- locking model can be validated by observing a correlation between
existent/absent mass accretion diagnosis and the star's rotational status as a slow/fast rotator.
Studies supporting the disk-locking scenario found that during the first few Myr of their
evolution, PMS stars of solar mass (0.4M$_\odot$$\le$M$\le$1.4M$_\odot$) present a bi-modal period
distribution, with disk-bearing stars having typical periods between 3-10 days, while non-disked
stars have periods between 1-7 days
\citep[e.g.][]{2004Rebull,Irwin2008+,RodriguezLedesma2009,Littlefair2010+, Affer2013,2016Venuti}.
Recently, \citet{2015VasconcelosBouvier} used semi-empirical Monte Carlo simulations to
investigate the effect of the disk-locking hypothesis on the period distributions of groups of
coeval stars.  They adopted a model where accreting stars have constant rotational periods
(disk-locking), and non-disked stars conserve AM. They applied to the model values of mass accretion
rates and disk lifetimes from the literature. They succeeded in reproducing the scenario observed in
young clusters of several ages when they started their simulations with a bimodal period
distribution at the age of 1 Myr: with disked stars rotating with periods of 8$\pm$6 days, and
non-disked stars rotating with periods of 3$\pm$2 days. Nevertheless, disk-locking seems to be less
efficient for very low mass stars \citep[][]{2005Lamm,BouvierMatt2013ReviewII}, and \citet{2010Cody}
found no correlation between disk presence and rotational periods for stars with masses below
0.5M$_\odot$.

Interpreting period distributions in the light of disk-locking models may be a delicate process.
Often, conclusions regarding the statistical significance of the differences between the rotational
period distributions for classical TTS \citep[CTTS][]{} and weak-lined TTS \citep[WTTS][]{} -
commonly seen as supportive of disk-locking scenario - are a controversial. The results reported in
some studies are often not verified by other studies, even for the same group of stars. This is
because external factors can easily introduce ambiguities in the period distribution interpretation.
Among the typical observational biases there are the 1 day aliasing phenomenon introduced by Earth's
rotation in ground-based observations (to be discussed in Section \ref{sec:periodic}), sample
incompleteness, periodic sample with small statistics numbers, and physical aspects like the fact
the rotational scenario is mass dependent, and therefore uncertainties in mass estimation can easily
contaminate the results. Another recurrent physical contamination factor is the disk-diagnosis used
to identify disk-bearing stars. Several studies (including the present one) used IR excess as
indicative of disk presence, and despite being a good diagnosis for a dusty local environment,
interpreted as a disk, it does not tell if there is indeed an active accretion process in the disk,
and thus whether the star is still magnetically interacting with the disk. 

Even with all those caveats, the best way to date to study AM evolution is still to measure rotational periods of groups of 
coeval stars in open cluster or associations, and then to assemble period distributions for clusters of different ages
in an evolutionary sequence. In this direction, thousands of rotational periods of stars in young clusters and 
associations have been measured during the last decades \citep[for a review see][]{BouvierMatt2013ReviewII}. 
However, the assumption that each group represents a piece of the same time-line assumes that global 
environmental conditions do not play a significant role in the overall rotational evolution. Notwithstanding,
some discrepant observational results for regions with similar age show that environmental conditions may have an
important effect in the evolution of AM during the first Myr. Some examples are: the case of CepOB3b region 
\citep[with age 4 - 5 Myr according to ][]{Littlefair2010+}, where the authors reported quite different rotational 
period distributions at very low mass (M$\leq$0.4M$_\odot$) than \citet{Irwin2008+} reported for the similar aged 
NGC2362; and the case of IC348 and NGC2264 \citep[age 1 - 3 Myr according to][]{Littlefair2005}, where stars of the
former cluster were reported to be rotating slower than those of the latter cluster.

It is, therefore, mandatory to build statistically significant samples of stats with measured
rotational periods in young regions with similar ages but different environments, in order to
improve our understanding of the environmental conditions role in regulating the AM during PMS. In
this context, massive young associations such as Cygnus OB2 (CygOB2) are valuable targets for
investigating the effects of environment on the AM evolution.

CygOB2 is the closest young massive association to the Sun. \citet{2015Kiminki} recently reviewed its
distance by studying four double-lined eclipsing binaries within the association, analysed using both photometry
and spectroscopy. They found an average distance of 1.33$\pm$0.06 kpc to the association.

CygOB2 massive population has been investigated by photometric and spectroscopic studies in both
optical and infrared bands
\citep[e.g.][]{Reddish66+,torres91+,knod2000,comeron2002+,hanson2003,Drew2008+,2015Rauw,2015WrightMassive,
2015Kiminki,2016WrightDance}, and it is known for harbouring some of the brightest stars in the
Galaxy. Some examples of its rich population are: the peculiar B supergiant CygOB2$\#$12
\citep{MassThomp91}, two O3If stars  \citep[$\#$7 and 22-A from][]{Walborn2002}, and some Wolf-Rayet
stars \citep[e.g., WR 142a from ][]{Pasquali2002+}. There are 169 confirmed OB stars
\citep{2015WrightMassive} among the association members. 

\citet{2016WrightDance}, as part of the DANCe \citep[Dynamical Analysis of Nearby
Clusters,][]{DANCe1st} project, used high-precision proper motions of stars in the association to
investigate its kinematic and dynamic.  They suggested that CygOB2 formed pretty much as it is
today: a highly sub-structured, globally unbound association.  Beyond its massive population, CygOB2
is also a valuable target for studying the environmental influence on the formation and early
evolution of low mass stars (M$\le1.4$M$_\odot$), and, in particular, to probe low mass star
evolution in the vicinity of massive stars.

In the last decade, CygOB2 low mass population has been the target of several studies
\citep[e.g.,][]{2008Vink,wrightdrake2009,Guarcello2012+,gdw13,Wright2014b+,gdw15,gdw16} and while
some massive stars have ages as low as 2 Myr \citep[e.g.,][]{hanson2003}, the low mass population
age ranges mainly between $\sim$3 Myr and $\sim$5 Myr \citep{wright2010+}. \citet{gdw13} used wide and
deep photometry from r band to 24 $\mu$m to unveil its disk-bearing population, finding 1843
disk-bearing candidates. As result of the Chandra Cygnus OB2 Legacy Survey, \citet{Wright2014b+}
identified 7924 X-ray sources in the direction of the association. \citet{gdw15} found that 5703 of
the X-ray sources had optical and/or infrared counterparts. Using the optical and infrared
photometry merged by \citet{gdw15}, \citet{2016Kashyap} classified 5022 sources as candidate
members. \citet{gdw16} recently used the disk-bearing and non-disked candidate members to
investigate the dissipation timescale of protoplanetary disks of low mass stars in the vicinity of
massive stars. They analysed the spatial variation of the disk fraction across the association, and
its correlation with the local ultraviolet radiation and stellar density. They found evidence that
disks are more rapidly dissipated in regions with higher stellar density and more intense UV
radiation within the association. They also found that disk dissipation due to close encounters is
negligible in the association, and that disk dissipation is dominated by photoevaporation. 

We used the results of a near-infrared (NIR) photometric variability survey in the direction of the
association to address some questions regarding the association's low mass population. We will
present the first results of the survey in two papers. In this first paper we focus on a sample of
periodic stars that were listed as candidate members in the literature. As most of periodic
variability can be explained by the rotational modulation by spots at the stellar surface, we
analyse the periodic sample in the context of angular momentum (AM) evolution for young stars. The
general characteristics of the NIR variable sample, and the description of their variable
behaviour inside the colour-magnitude and colour-colour diagrams will be presented in the second
paper (Roquette et al 2017, in preparation). 

The present paper is organized as follows: In Section \ref{sec:sec2} we outline the observations
carried for the present survey, the data processing, and the light curve production, as well as data
taken from the literature used to complement the study. In Section \ref{sec:sec3} we describe the
methods applied for time series analysis and periodicity search. In Section \ref{sec:sec4} we
present the results regarding the periodic stars, and in Section \ref{sec:sec5} we present
conclusions and discuss the results inside the general young stellar AM scenario.

\section{Analyzed Data}
\label{sec:sec2}
\subsection{Observations}
\begin{figure}[tb]
   \includegraphics[width=\columnwidth]{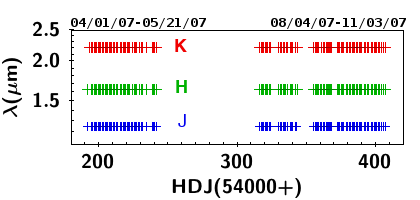}
 \caption{\label{fig:dates}Scheme showing the nights when observations were taken in the given filters.}
 \end{figure}

Our observational dataset was obtained with the 3.8m United Kingdom Infra-Red Telescope (UKIRT), at
Manua Kea, Hawaii, equipped with the Wide Field Camera \citep[WFCAM][]{wfcam},  programs U/07A/H16
and U/07B/H60. Our complete dataset is composed by up to 115 nights observed using the J, H and K
filter \citep{filtroukirt}. The observations were carried during 2007 in two seasons (Figure
\ref{fig:dates}): The first season comprises 43 observed nights between April 1st, and May 21th; The
second season comprises 73 observed nights between August 4th and November 3nd.  The two
observational seasons span a total of 217 days. The exposures were short, 2 seconds in each
filter.
 
\begin{figure}[tb]
  \centering
  \includegraphics[width=\columnwidth]{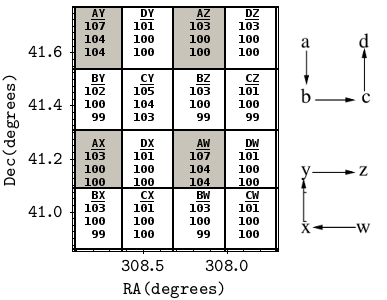}
\caption{\label{fig:wfcam}Schematic representation of WFCAM layout. A sequence of 4 exposures (ABCD) 
  with the 4 CCDs (WXYZ) produces a mosaic of 16 observed regions covering a total area of 0.87 
  squared degrees in the sky. For each region composing the mosaic, we show also the number of 
  observations in J, H, and K filter after data processing and cleaning.}
\end{figure}

The WFCAM is composed of four 2048 $\times$ 2048 Rockwell Hawaii-II detectors \citep[][hereafter
CCDs W, X, Y, and Z]{wfcam}. The detectors are spaced with a separation of 94$\%$ of each detector's
width, such that four exposures (exposures A, B, C and D) are required in order to image a
contiguous area of 0.87 squared degrees.  WFCAM's layout is schematically shown in Figure
\ref{fig:wfcam}. The observed area was centered on $\alpha_{2000}=20^h33^m$,
$\delta_{2000}=+41^o12'$, which comprises approximately the center of CygOB2 association.

The data were pipeline-reduced and calibrated at the Cambridge Astronomy Survey Unit \citep[CASU; 
Irwin et al.  2004;][]{wfcamcalibration}, and a source catalogue was provided. The source catalogue 
is composed by a set of fluxes measured with different aperture radii per source. As we need a 
single number for the flux in all images (in order to link the objects building consistent light 
curves), we followed CASU's documentation and adopted flux number 3, which is a soft-edge aperture 
of $1\farcs 0$ radius. Data calibration was made by CASU's pipeline using 2MASS sources with 
extinction-corrected colour 0.0$\le$J-K$\le$1.0 and signal-to-noise ratio $\ge 10$ in each filter 
\citep{wfcamcalibration}. 

The images and catalogues were retrieved from the CASU server in January, 2008. Due to WFCAM 
4-detectors layout (Figure \ref{fig:wfcam}), one night of observation in J, H, and K filters is 
composed by 4$\times$4$\times$3=48 frames/source catalogues. In order to build a light-curve 
catalogue from the 5640 tables, we created an IDL procedure to read, manipulate, and link sources 
from the tables provided by CASU. We excluded all sources with the Classification Flag in CASU's 
catalogues indicating noise, borderline stellar or saturated objects. 

\paragraph{Candidate Member Catalogue:} Our main goal is to determine the variability 
characteristics of young stellar objects (YSO) that belong to CygOB2 association. We searched the 
literature for candidate members  and used their coordinates to build an input catalogue for 
cross-matching and merging all CASU tables together in a single multi-band light curve catalogue.  

The candidate member catalogues used to compose the input catalogue were: 
 \begin{itemize}
   \item{} The list of disk-bearing stars from \citet[][, hereafter GDW13]{gdw13}, which contains
     1843 stars;
  \item{}The X-ray sources from Chandra Cygnus OB2 Legacy Survey with optical/infrared counterparts 
    from \citet[][hereafter GDW15]{gdw15,gdw16}, considered as members by Kashyap (2017, in 
    preparation): 4864 sources.
 \end{itemize}

Each object from this input catalogue was examined for companions in CASU tables inside a radius of 
$0\farcs 75$.  Every time two objects were found inside this search radius, both objects were 
excluded in order to avoid contaminations into the light curves, and this selection rule accounts 
for most objects we could not recover inside our field of view (FOV). The catalogue was built 
processing region by region (AW, AX, and so on, cf. Figure \ref{fig:wfcam}). At the end, the 16 
light curve catalogues were merged and objects in the overlapping region between the CCDs were 
merged into a single identifier. 
 
\begin{figure}[tb]
\includegraphics[width=\columnwidth]{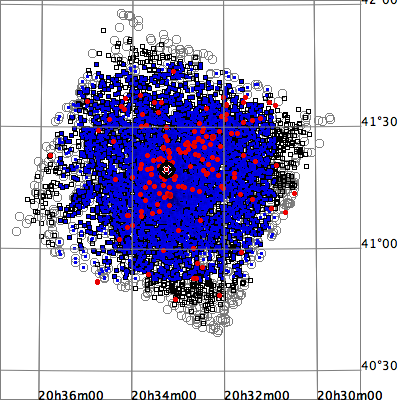}
\caption{\label{fig:literature} Spatial distribution of candidate members. Stars from GDW15 are 
shown as black squares, and stars from GDW13 are shown as gray points. Membership candidates with 
valid light curve in the present study are shown as blue dots. OB stars from 
\citet{2015WrightMassive} are shown as red circles.}
\end{figure}

There were 100 OB stars on GDW15 and all those stars were saturated in our images. GDW15 and GDW13 
have 403 low mass stars in common, and we recovered 354 of them (87.84$\%$). We recovered 1272 
(69.02$\%$) disk-bearing stars from GDW13 and 4165 (85.63$\%$) X-ray sources from  GDW15. Our final 
candidate member light-curve catalogue is composed by 5083 stars. The spatial distribution of those 
stars is shown in Figure \ref{fig:literature}.
 
\paragraph{Control Sample:} We built a control sample including all stars in the three central 
regions (AW, BZ and CY in Figure \ref{fig:wfcam}) regardless of confirmed membership. Non-variable 
stars selected from the control sample (\textbf{as described in Section \ref{sec:simulations}}) were 
used to estimate limit values of the statistical indexes used for evaluating periodic stars, such as 
the false alarm probability for peaks in the Lomb/Scargle periodogram as it will be described on 
Section \ref{sec:periodic}. For each region, we choose the best night in K band in terms of seeing, 
and generated an initial catalogue with the coordinates of all the stars observed on that night. The 
typical number of detected sources in the central fields was around 25000 for nights with good 
seeing. For each subsequent night and/or filter exposure for that region, we used a searching radius 
of 0.75 arcsec for each object already in the catalog.  After a complete inspection, the objects in 
each exposure not matched with the light curve catalogue were added to it as new objects. A 
human-operator was necessary in the procedure in order to judge parameters and make decisions in how 
to proceed in case of ambiguities due to:  1- objects in the central fields being too crowded and, 
causing more than one object to be found inside the 0.75 arcsec radius search; 2- Seeing variations 
from night to night, causing variations in how many resolved objects were detected from one night to 
the other. 3- some objects presented proper motion with varying coordinates. 4- spurious objects 
inside the source catalogue, caused by defects in the data reduction which were not flagged by CASU 
processing. 

\begin{figure}[tb]
  \centering
  \includegraphics[width=\columnwidth]{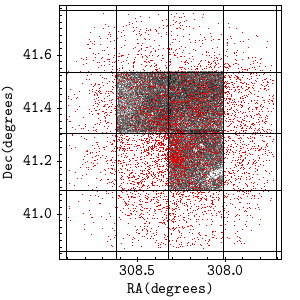}
  \caption{\label{fig:catalogue}Spatial Distribution of the stars with a analysed light curve. 
  Control Sample objects are shown as gray circles, and candidate member stars are shown as red 
dots. }
\end{figure}

Using this method, we catalogued 42777 objects with more than 10 valid observations in at least one 
filter.  Objects catalogued in the control sample are shown in gray in Figure \ref{fig:catalogue}, 
while objects in the candidate member catalogue are shown as red dots.

\begin{figure}[tb]
  \centering
  \includegraphics[width=0.48\columnwidth]{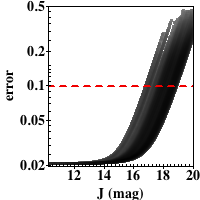}\includegraphics[width=0.48\columnwidth]{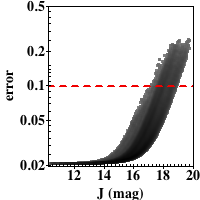}\\
  \includegraphics[width=0.48\columnwidth]{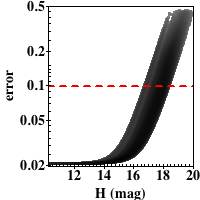}\includegraphics[width=0.48\columnwidth]{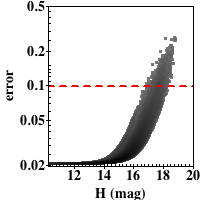}\\
  \includegraphics[width=0.48\columnwidth]{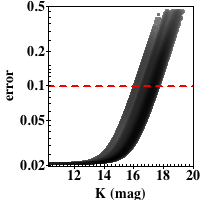}\includegraphics[width=0.48\columnwidth]{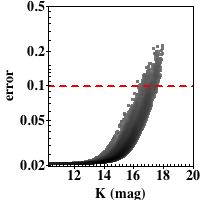}\\
  \caption{\label{fig:error}Errors versus magnitude distributions for all data-points in both 
  control sample catalogue (left) and candidate member catalogue (right), for J, H, and K filters. 
The limit of 0.1 magnitudes for the error, adopted in the present study, is shown as a red dashed 
line.}
\end{figure}

\paragraph{Corrected Errors:} After completing the multi-band light curve catalogues, we applied the 
empirically derived correction presented by \citet{wfcamcalibration} to the pipeline-estimated 
photometric errors.

\begin{equation}
M^2=cE^2+s^2,
\label{eq:error}
\end{equation}

\noindent where $M$ is the correct measured total error, $E$ is the pipeline-estimated photometric 
error, and $c$ (=1.082), and $s$ (=0.021) were empirically determined by \citet{wfcamcalibration}. 
With this updated error, we confirmed the 2$\%$ level of night to night variations estimated by 
\citet{wfcamcalibration} for UKIRT/WFCAM data pipe-line reduced by the CASU, as can be verified in 
Figure \ref{fig:error}.

For each light curve, in order to remove individual points with error unusually higher than the 
light curve's mean error, we did a two iterations 2$\sigma$-clipping in the error distribution 
around the mean error. After cleaning the complete candidate member sample, we analysed the outlier 
points and identified and removed the nights suffering from systematic errors (those for which more 
than 40$\%$ of the valid points in a certain filter).  Eight observed nights were removed from the 
light curves given this criterion. Figure \ref{fig:error} shows the error distribution for each 
filter after removing these points and nights. To guarantee high quality photometry, we only used 
data points with error smaller than 0.1 magnitude. 

\subsection{Data from Literature}
\label{sec:literature}

GDW13 composed an Optical-Infrared (OIR) catalogue using wide and deep photometry, from r band to 
24 $\mu$m, extracted from the literature. Their OIR catalogue is composed by 328540 sources in the 
field of CygOB2. The surveys used by the authors to build their catalogue were: 
\begin{description}
  \item{GTC/OSIRIS Catalogue \citep[hereafter GDW12;][]{Guarcello2012+}}. r, i and z bands, 65349 
    sources.  For sources with good photometry, as defined by GDW12, the catalog reaches r = 
    $25^\mathrm{m}$; for objects at the distance of 1400 pc, and using a 3.5 Myr isochrone 
    \citep{wright2010+}, with the average extinction $\mathrm{A}_V=4\fm 3$ (from GDW12), this limit 
    corresponds to a 0.16 $\msun$ star \citep{gdw13}.

  \item{WFC/IPHAS catalogue in} r' , i' , and H$\alpha$ filters, 24072 sources. First release 
    \citep{IPHAS2005} for GDW13 and second release \citep{IPHAS2014} for GDW15. The limit for good 
    photometry is around $\mathrm{r}^{\prime}$=$21\fm 5$ \citep{gdw13}. 

\item{SDSS DR8 catalogue \citep{SDSSDR8}}. u, g, r, i, and z bands, 27531 sources. The limit for 
  good photometry is at r=22$^m$ ($\sim$0.4 $\msun$) but with a higher limit of saturation than 
  GTC/OSIRIS catalogue allowing the study of stars up to r=16$^\mathrm{m}$ 
  ($\sim$2.5$\mathrm{M}_{\sun}$ \citep[$\sim 2.5M_\odot$][]{gdw13}.  

\item{UKIDSS/GPS Catalogue} J, H, and K bands, 273473 sources. The original UKIDSS 
  \citep{UKIDSSGPS2008} photometry extraction was redone by \citet{gdw13}, and reaches J=21$^m$ 
  J=$21^\mathrm{m}$ corresponding to a mass limit of $\sim$0.1 $\msun$ at the distance of Cygnus 
  OB2.

\item{2MASS/PSC \citep{2MASS}.} J, H, and K$_s$ filters, 43485 sources. According to \citet{gdw13}, 
  it was used because it has a higher limit of saturation than UKIDSS.

\item{Spitzer Legacy Survey of the Cygnus X region \citep{SpitzerCygX}.}  IRAC 3.6, 4.5, 5.8, 8.0 
  $\mu$m, and MIPS 24 $\mu$m (149381 sources) bands with sources detected down to 0.5 $\msun$ 
  \citep{gdw13}.
\end{description}

To select the disk-bearing stars, GDW13 applied several disk-diagnosis techniques to their
OIR-catalogue, finding 1843 candidates. They also inferred the evolutionary status of the disked
objects by studying their infrared spectral index, $\alpha$=$\frac{\mathrm{d}\log(\lambda
F_\lambda)}{\mathrm{d}\log(\lambda)}$, using \citet{Wilking2001+} classification scheme: 8.4$\%$ of
the disk-bearing stars were class I, 13.1$\%$ were flat-spectrum sources, 72.9$\%$ were class II
sources, 2.3$\%$ were pre-transition disks, and 3.3$\%$ were transitional disks.

As part of the Chandra Cygnus OB2 Legacy Survey \citep{Wright2014b+}, a region of 1 square degree
was covered using 36 Chandra/ACIS-I overlapping fields, and the authors detected and verified 7924
X-ray point sources in the observed region. GDW15 cross matched those X-ray sources with GDW13 OIR
catalogue and found 5703 X-ray sources with OIR counterpart. From the list of X-ray OIR sources,
\citet{2016Kashyap} selected 4864 candidate members.

\subsection{\textbf{Completeness Of The Analysed Data}}
\label{sec:completeness}
\begin{figure}[tb]
  \centering
  \includegraphics[width=0.48\columnwidth]{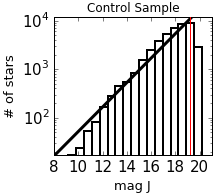}
  \includegraphics[width=0.48\columnwidth]{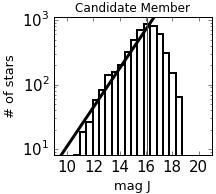}
  \includegraphics[width=0.48\columnwidth]{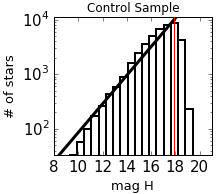}
  \includegraphics[width=0.48\columnwidth]{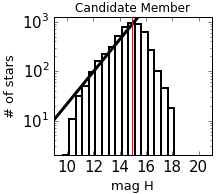}
  \includegraphics[width=0.48\columnwidth]{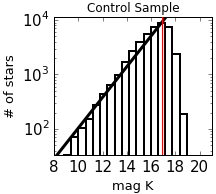}
  \includegraphics[width=0.48\columnwidth]{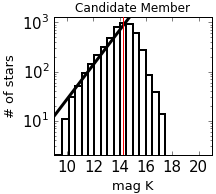}
  \caption{\label{fig:hmag}Magnitude distributions for the control sample (left) and candidate
    member sample (right), for J, H, and K filters.}
\end{figure}

\textbf{
The completeness of the analysed data was estimated based on the drop in the number of detected objects
in the magnitude distribution histograms for J, H, and K filters, which are shown in Figure
\ref{fig:hmag} for the control and member candidate samples.}
\textbf{
  For the control sample (left plots), the limit magnitude detected in
  each band was J$_\mathrm{max}\sim$ 20.2 mag, H$_\mathrm{max}\sim$19.4 mag, and
  K$_\mathrm{max}\sim$19.0 mag. The samples are complete up to magnitudes
  J$_\mathrm{complete}\sim$ 19.2 mag, H$_\mathrm{complete}\sim$17.9 mag, and
  K$_\mathrm{complete}\sim$16.9 mag.
}
\textbf{
For the candidate member sample (right plots), the limit magnitude detected
in each band was J$_\mathrm{max}\sim$ 18.8 mag, H$_\mathrm{max}\sim$18.2 mag, and
K$_\mathrm{max}\sim$17.4 mag. The samples are complete up to magnitudes
J$_\mathrm{complete}\sim$ 16.1 mag, H$_\mathrm{complete}\sim$14.9 mag, and
K$_\mathrm{complete}\sim$14.3 mag.}

\textbf{
From Figure \ref{fig:hmag} one can see that the control sample goes deeper than the candidate member
sample. This happens due to the lower completeness limits in the studies used to compose the
candidate member sample. Also, there can be an incompleteness towards the brighter stars in both
samples, but this incompleteness is not thought to interfere on the present work, as at the
distance of CygOB2, the brightest stars in our sample would correspond to stars with intermediate
mass, and those are out of the mass range of interest in the present work.}

\section{Time-Series Analysis} 
\label{sec:sec3}

Initial variable selection was done using the Stetson Variability Index \citep{Stet1996}, defined 
as:

\begin{equation}
S=\frac{\sum_{i=1}^p w_i \mathrm{sgn}(P_i)\sqrt{|P_i|}}{\sum_{j=1}^p w_i}
\end{equation}

\noindent where $i$ is a pair of observations, which has a weight w$_i$, and $p$ is the total number 
of pairs of observations. $P_i$ is defined as the product of the normalized residuals of two 
observations $j$ and $k$ that constitute the $i$-th pair, $P_i$ = $\delta_{j(i)}\delta_{k(i)}$, and 
as $P_i$=$\delta^2_{j(i)}-1$ when there is only one valid observation ($j=k$). If one night has 
valid observations in all J, H and K filters, then there are three pairs of observations for that 
night. Following \citet{2001Carpenter}, if the star has valid J, H, and K, then each pair of 
observation has weight $w_k$=$\frac{2}{3}$ (total weight of 2 for the whole set), when there was a 
missing point, i. e., nonexistent observation in one or two filters, then a weight $w_k$=1 was 
assigned. The normalized residuals are defined as:

\begin{equation}
\delta_k=\sqrt{\frac{n}{n-1}}\frac{m_k-\bar{m}}{\sigma_k},
\end{equation}

\noindent for a given filter, where $n$ is the number of measurements used for determining the mean 
magnitude, $\bar{m}$, and $\sigma_k$ is the photometric uncertainty for the  measurement $k$. 

\begin{figure}[tb]
   \includegraphics[width=0.5\columnwidth]{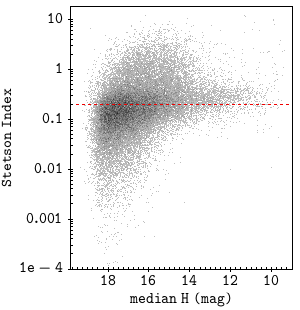}\\
   \includegraphics[width=0.5\columnwidth]{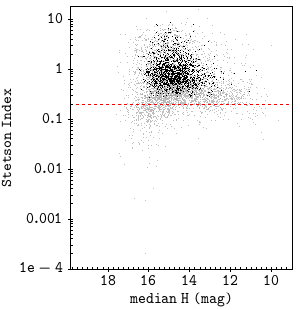}\\
   \caption{\label{fig:stetson} Distribution of 3-band Stetson variability index versus mean H
     magnitude. Top: 36365 objects from control sample with 3 band light curve. Higher density of
     points is shown as darker gray. Bottom: 5149 stars from the candidate member catalogue. In
     both panels, the 3-band light curve is plotted as light gray dots, and darker gray shows
     regions with high density of points. The dashed red lines show the selection value $S=0.25$. In
   the right plot, periodic stars are plotted in black.}
 \end{figure}

Designed this way, the Stetson Index sets uncorrelated non-variable stars with values of $S\sim0$, 
and significant variables with $S\ge1$. Different authors adopt different Stetson index limits for 
accounting for low-amplitude variable stars. For example, \citet{2001Carpenter} adopts $S\ge0.55$, 
and \citet{2008Plavchan} found periodic variables down to $S\sim0.2$. For the purposes of the 
present work we adopted a selection limit of $S\ge0.25$for variable stars. The distribution of 
Stetson Variability Index as a function of magnitude H is shown in Figure \ref{fig:stetson}.

 From the candidate member sample, 4079 (80$\%$) stars met the Stetson Variability criterion. We did 
 a visual inspection of these candidate variable stars and organized them in morphological 
 classes\footnote{ Details on the morphological classification will be present on \citet[in 
   preparation][]{2017Roquette}, but for the purposes of the present study it is worth mentioning 
   that: eclipse-like variables are stars which light curve points are most of the time in a maximum 
   value of bright, but present several dimmings in bright (They may also be periodic). Candidate 
   periodic stars present light curve oscillating between maximum and minimum values. Other 
   variables encompasses all stars with a visibly variable light curves but that did not fit in the 
   previous categories, like stars with long term variability (e.g., stars that are slowly dimming 
   with time), and also stars with a mixed light curve (e.g. long term variability mixed with a 
 short term variability in their light curves). Stars that met the Stetson Index criteria, but 
 presented low amplitude stochastic variability in their light curves were not considered in the 
 study.}: eclipse-like variables (110 objects, 2.2$\%$ of the total candidate member sample), 
 candidate periodic variables (1679, 33.0$\%$), and non-periodic variable stars (1288, 25.3$\%$). 
 1002 stars (20$\%$) met the Stetson variability index criteria, but were not considered as 
 variables after light curve visual inspection. In the present study, since we target measurements 
 of rotational periods, we will focus only on the first group. Stars from the candidate periodic 
 sample with confirmed period (Section \ref{sec:periodic}) are shown in black in the right panel of 
 Figure \ref{fig:stetson}.

\subsection{Period Search}
\label{sec:periodic}

The main technique used for identifying and determining periods was the Lomb Scargle Normalized
Periodogram \citep[hereafter LSP][]{lomb,scargle2}, a widely used modified version of the classical
periodogram based on fast Fourier transforms, that can be applied to unevenly spaced datasets. The
LSP algorithm used here was implemented according to \citet{numericalr} and \citet{horne}, and it
was normalized by the total variance of the data. This normalization guarantees that the power of a
certain frequency's peak in the periodogram, ($z$), is related to the  false alarm probability (FAP)
for that frequency as   
\begin{equation}
  \label{eq:FAPh}
  \mathrm{FAP}=1-(1-e^{-z})^{N_i},
\end{equation}
were $N_i$ is the number of independent frequencies used to compose the periodogram.

To apply the LSP algorithm to the sample, an oversampling factor of 260, and a scale factor of 5 for 
the Nyquist Frequency were used, a choice that sets the lower limit in the period search as 
$\sim$0.83 day and that will be discussed in more details in Section \ref{sec:contamination}. For 
evenly spaced data, the higher limit in the frequency search would be given by 
$f_\mathrm{Nyquist}=\frac{1}{2\Delta t}$, were $\Delta t$ is the time between consecutive data 
points. For unevenly spaced data, the Nyquist frequency, calculated with $\Delta t$ being the mean 
time between two consecutive measurements, gives only a rough estimation for the higher limit in the 
frequency search \citep{2004Scholz}. Since we are dealing with unevenly spaced data, we extended our 
search over higher frequencies than the limit imposed by the Nyquist frequency (which is around 2 
days for our dataset), and potential contamination due this choice will be discussed on Section 
\ref{sec:simulations}. Our dataset is composed of two observational seasons of $\sim$45 days and 
$\sim$75 days, respectively. For the lower limit in the frequency search, we adopted the resolution 
of the smaller season, i.e. $\frac{1}{45\;\mathrm{days}}$. Hence, periods were searched in the 
interval 0.83-45 days.

The search for periods via LSP consists in studying the highest peaks inside the periodogram, and
determining their significance. Judging a certain frequency's power peak significance can be a
tricky task and it is until nowadays one of the main limitations in studies regarding periodic
stars. First, there is the 1 day aliasing phenomenon that can be written as:

\begin{equation}
  \label{eq:alias}
  P_\mathrm{measured}=\frac{1}{n\pm\frac{1}P_\mathrm{True}}
\end{equation}

\noindent with $n$ integer, and it is caused by the limitations in observation imposed by the
Earth's rotation \citep[cf.,][]{48Tanner}. Second, we are here dealing with young stars and they
typically show irregular variability that may be mixed with the periodic signal. The combination of
irregular variability with discrete and uneven sampling may cause the occurrence of spurious
periodogram power peaks, that despite being high peaks, are not truly associated to a periodic
signal. 

\citet{horne} reduced the problem of estimating the FAP (FAP$_h$ when referring to their concept of
FAP) to the problem of finding the number of independent frequencies adequate to be applied in
Equation \ref{eq:FAPh}.  They used Monte Carlo simulations to generate a large number of data sets
with pseudo-Gaussian noise, and different time-samplings. From the simulated data, they estimated
$N_i$ for a set of unevenly and non clumped data, as $N_i$=$-6.3$$+$$1.2N$$+$$0.00098N^2$, where $N$
is the number of valid data points. Despite being largely used in the literature, this method for
estimating FAP$_h$ may not be adequate. As pointed out by \citet{Littlefair2010+}, FAP$_h$
calculations via Monte Carlo simulations with gaussian noise are not reliable, since they can only
account for variability due to photometric errors, while often there is also some intrinsic
variability characteristic of young stellar objects. \citet{Littlefair2010+} propose, as an
alternative, to use the light curves themselves as a mean of estimating the height of spurious peaks
due their intrinsic variability.  When using the light curves themselves, we are accounting for
spurious peaks introduced into the periodogram by all factors affecting the dataset, from imperfect
photometry to intrinsic variability \citep{Littlefair2010+}. 

\paragraph{False Alarm Probability from a Control Sample:}\label{sec:FAP} When studying the sample of member
candidates of CygOB2, we expect to be dealing with a sample of young stars, and possibly with a high fraction
of periodic stars. On the contrary, when studying the sample of all objects in CygOB2 FOV, we expect a sample
rich in field stars, and a lower fraction of periodic stars. This gives us an alternative method for estimating
the FAP, that consists in studying known constant stars in our field of view and estimating the recurrence and
typical height of spurious periodogram peaks for them.

\begin{figure}[tb]
  \centering
  \includegraphics[width=0.5\columnwidth]{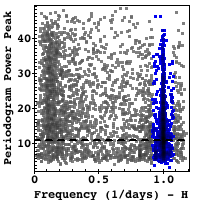}\includegraphics[width=0.5\columnwidth]{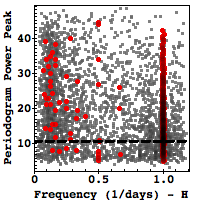}\\
  \includegraphics[width=0.5\columnwidth]{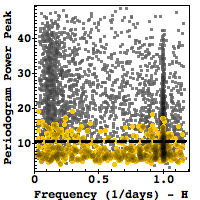}\includegraphics[width=0.5\columnwidth]{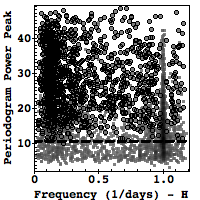}\\
  \caption{\label{fig:PPfreq} Periodogram Power Peak versus frequency from the selected peak
    distributions for H-band. All candidate members are shown as gray dots. (Top-left) Stars
    discarded due frequencies around 1 day$^{-1}$ are shown in blue. (Top-right) Stars discarded due
    to S-statistics limit are shown in red.  (Bottom-left) Stars discarded due RL-statistics limit
  are shown in yellow. (Bottom-right) Stars selected as periodic stars are shown as black empty
circles. The PFA$_c$ limit for H-band is shown as a black dashed line. }
\end{figure}

We estimated the FAP by using data from the control sample. LSPs were calculated for all the objects
in the sample, and the power of the highest periodogram peak for each object was recorded. A sample
of objects with frequency of the highest peak in the range $0.3<f<0.5\mathrm{day}^{-1}$\footnote{We
  chose this interval based on the frequency vs. periodogram power peak plots in Figure
\ref{fig:PPfreq}: this frequency interval is outside the bulk produced by the 1 day alias, and also
outside the bulk of longer periods (lower frequencies).}, and Stetson Variability Index smaller than
$S=0.15$ was selected, building this way a constant star sample, composed by 3999, 3077 and 3064
objects in J, H and K filters respectively. \textbf{The false alarm probability, FAP$_c$, was
  estimated from the cumulative distribution of the highest periodogram power peak in each filter as
the power peak bin that contains the desired percentage of constant stars data.} This way, the
  0.01$\%$ FAP$_c$ level for J, H, and K filters were found at the power value 11.11, 10.52 and
  10.47 respectively.

In spite of being a more accurate FAP estimator than analytical estimates or Monte Carlos
simulations as derived by \citet{horne}, FAP$_c$ gives only a reference value. Using it alone as a
cut for selecting periodic stars may minimize spurious detections, but will not be enough for
rejecting all of them. As discussed by \citet{Littlefair2010+}, very irregular variables and stars
with poor rotational phase coverage are two common sources of contaminants in such analysis. Looking
for a sample with the least contamination possible, we relied on the comparison of the periodograms
in each of the three observed bands, and in two complementary statistical analysis, independent on
the LSP.

\paragraph{Automatized Period Search:} The first test for periodicity was to evaluate the existence
of significant peaks in each of the J, H, and K periodograms. In some cases, the same frequency was
responsible for the highest peak in all the three periodograms, and thus that frequency was recorded
as a possible valid periodic signal. Due to some missing points, to differences in photometric
errors and in amplitudes, the same periodic signal sometimes produced different values of power peak
for each band. In particular, some missing points may favour the detection of an alias instead of
the true period. \textbf{Therefore, when the highest peak frequency was not the same for each of the
three periodograms, we chose the one with the highest power.} We then checked in the other two bands
periodograms if the same frequency was responsible for one of the three highest peaks. If it was the
case, that frequency and its power peak in each filter was recorded. If the same frequency was not
found within the three highest peak in each filter's periodogram, or if its power was smaller than
the FAP$_c$ of 0.01$\%$, the light curve was rejected as periodic. Periodogram Power Peak versus
frequency from the selected peaks are shown in Figure \ref{fig:PPfreq} for the H-band. All candidate
members with valid periodograms are shown as gray dots.

Next, we used the String/Rope length method \citep{Clarke2002,65LK}, which is a non-parametric
period search method based on Lafler-Kinman statistics \citep{65LK}. For each trial period, the
original data is folded in phase and re-ordered for ascending phases, and the String-Length
statistics (hereafter SL-statistics) was calculated as the summation of the squares of the
differences between the consecutive phases re-ordered measurement values, normalized by the data's
variance. Given the normalization proposed by \citet{Clarke2002}, values of SL-statistics will
fluctuate around the unity, with periodic stars presenting minimum values.
The SL-statistics can be extended to the case of multi-band data, that \citet{Clarke2002} calls the
Rope-Length statistics (RL-statistics). In the multi-band case, the SL-statistics of each band is
summed and divided by the total number of bands. SL and RL-statistics values can be calculated for a
set of trial periods, and used to compose a periodogram. In the present study we do not used
SL/RL-statistics for searching for periods, \textbf{as the two indexes were only evaluated for the
periods selected via LSP, and used for checking each period reliability}. We chose the limit value
for RL-statistics using Monte Carlo simulations with 78000 synthetic periodic light curves built as
described in Section \ref{sec:simulations}. For each synthetic light curve, RL-statistics was
calculated for the true period, and for the two aliases formed around 1 day. From the comparison
between the RL-statistics versus amplitude distributions for true period and aliases, we adopted an
RL-statistics of 0.8 as a limit between true periods and probable aliases.

The Saunders statistics \citep[S-statistics,][]{2006Saunders} is a technique that may be used for
investigating the aliasing effect of samplings. It is a normalized phase coverage metric and it is
defined as the sum of the squares of the distances between two consecutive points in the folded in
phase light curve, ordered for ascending phases, normalised by the value of the sum for an ideal
spacing of equally spaced observations across the phase space. An uniform phase coverage gives a
S-statistics of order unity. Growing irregular phase coverage makes S-statistics increase. The
S-statistics is especially good for removing the spurious periods arising from the 1 day$^{-1}$
natural frequency introduced by the Earth's rotation \citep[e.g.][]{Littlefair2010+}.  Monte Carlo
simulations were also done in order to obtain a limit value for S-statistics. For each synthetic
light curve in the simulated sample, LSP analysis was applied, and the highest peak in the
periodogram was recorded. Light curves for which the difference between the LSP period and the input
period was smaller than 10$\%$ the input period were selected, resulting in a subsample of 64102
synthetic light curves for which S-statistics was calculated. A cumulative distribution was built
for this subsample and from the bin from which 99$\%$ of the data is contained, we defined the limit
value of S-statistics equal to 5 for selecting stars with true period. 

The effect of the chosen RL-statistics and S-statistics limit selection are shown in the frequency
versus highest periodogram power peak distribution in Figure \ref{fig:PPfreq}. Even though the
filter for S-statistics significantly reduces the $\sim$1 day excess in the power peak versus
frequency distribution (cf. Figure \ref{fig:PPfreq}), it is not enough to completely account for
such aliasing effect. This conclusion was achieved after visually inspecting some light curves and
folded light curves, for stars with detected period very close to 1 day that were not filtered out
by S-statistics limits: in their majority, those stars were long term non periodic variables, and
not truly periodic stars. To deal with the remaining contamination, a filter for frequencies in the
range 0.92-1.08day$^{-1}$ was added. The choice of this range around 1 day was done based on
simulations discussed on \textbf{Appendix \ref{sec:contamination}.}

The final selection was composed by 1291 stars: 25 eclipse-like, 1256 form candidate periodic
variables list, 2 non-periodic variables (stars selected as periodic stars, but with no visible
periodic signal in the light curve), and 8 non-variable stars (stars not considered as variable
according to the visual selection). Each period was measured as the mean value of the inverse of the
frequencies obtained via LSP, $P=\frac{1}{f}$ for each filter. From that, it follows that the rms
error for each band period is $\delta P=\delta f\times P^2$. Simulations as described in Section
\ref{sec:simulations} were used for estimating the resolution $\delta f$ for each filter. For each
synthetic periodic light curve, LSP was calculated, and the highest periodogram power peak
frequency's full width at half maximum, FWHM, was estimated. We estimated $\delta f=<\mathrm{FWHM}>$
from the FWHM versus period distributions for each filter, and by propagating the error it follows
that the error in the periods are around $\delta P=0.002P^2$, which gives a 4.8 hours error for a 10
day period. 

\subsection{Period Detection Completeness and Reliability}
\label{sec:simulations}

Monte Carlo simulations were ran on synthetic light-curves in order to estimate the efficiency and
limitations of the techniques employed in the period search method. We used a modified version of
this type of simulation commonly presented in the literature
\citep[e.g.,][]{Littlefair2010+,2013Moraux}, since we used candidate members for building synthetic
light curves, instead of non-members in the field. To build synthetic light curves, ``constant''
stars were selected from the candidate members catalog as follows: First, stars with Stetson
variability index smaller than 0.15, and more than 60 valid observations in each filter were
selected. From those, we selected stars classified as non-variable stars during the visual light
curve inspection (stars with very small and uncorrelated variability). Finally, for each filter, we
analysed the peak to peak (ptp) amplitude versus magnitude distribution: using a bin of size one for
magnitudes between 10 and 18, we randomly selected stars in the minimum outlier of the distribution.
Whenever possible, two stars were selected per magnitude bin. A total of 39 stars were selected this
way, and as they were considered the most constant stars in the candidate member sample, their light
curves reflect the fingerprints of the dataset, including the window of observations and the kind of
photometric and instrumental error contained on the light curves. Each light curve in this constant
stars sample was then used as a template for building synthetic light curves. For each light-curve,
the time sampling, magnitude and error values were kept, and a periodic signal with random waveform,
period (P$_{in}$), amplitude, and phase was added to it. We generated 2000 synthetic light curves
per constant star, totalizing 78000 synthetic light curves. Uniform distributions were used to
generate the random periods, amplitudes and phases.

From the visual inspection phase in the light curve analysis, it was noticed that even though
periodic behaviour was quite often observed, they were usually not perfect sinusoidal signals. This
may have direct consequences in the period analysis performed, as certain waveforms may favour the
detection of aliases or harmonics in the LSP, instead of the true frequency. In fact, when inserting
periodic eclipse-like waveforms in our synthetic light curves, we notice a favouring for the
detection of harmonics, instead of the true period. As we already distinguished between eclipse-like
and candidate periodic light curves, eclipse-like waveforms will not be included in the completeness
analysis. For the simulations purposes, we used single period sine-wave, sawtooth wave, triangle
wave, and cycloidal wave as waveforms. The waveform was randomly chosen at the beginning of the
creation of each synthetic light curve. 

\begin{figure}[tb]
  \centering
  \includegraphics[width=0.95\columnwidth]{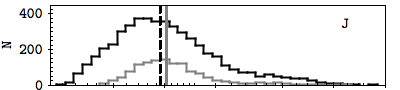}\\
  \includegraphics[width=0.95\columnwidth]{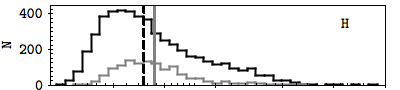}\\
  \includegraphics[width=0.95\columnwidth]{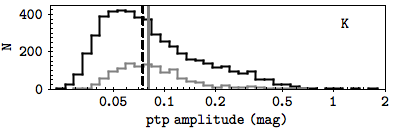}\\
  \caption{\label{fig:amplitude}Ptp amplitude distributions for each band. Black: Membership candidate catalogue.
    Gray: Periodic Stars (Section \ref{sec:periodic}). Black dashed lines show the median value for the member 
    candidate sample: 0.094, 0.075, and 0.074 mag for J, H, and K band respectively. Gray lines show the same 
    for the periodic sample: 0.102, 0.087, and 0.080 mag.}
\end{figure}

Two series of simulations were performed. The first one considers periods as large as the maximum
period searched (45 days, c.f., Section \ref{sec:periodic}), and as short as 0.2 days, which is
about the minimum rotational periods verified in other young regions
\citep[e.g.,][]{Irwin2008+,Littlefair2010+,2013Moraux}.  It also considers a large range of
amplitudes from 0.015-1.5 magnitudes. The second one considers amplitudes in the range 0.015-0.2
mag, and periods in the range 0.2-20.0 days, which considers a more realistic amplitude and period
upper limit given that Figures \ref{fig:amplitude} and \ref{fig:distP} shows that most of the
stars have amplitude smaller than 0.2 magnitudes, and period smaller than 20 days. Random ptp
amplitudes were generated for the J filter, and H and K magnitudes were set according to the ratio
between the median value for each band periodic star ptp amplitude distribution (gray dashed lines,
in Figure \ref{fig:amplitude}): $\frac{A_J}{A_H}=1.17$, and  $\frac{A_J}{A_K}=1.28$.

\begin{figure}[bt]
  \centering
  \includegraphics[width=0.33\columnwidth]{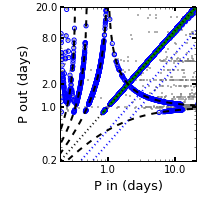}\includegraphics[width=0.33\columnwidth]{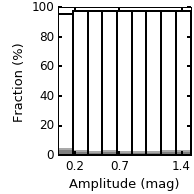}\includegraphics[width=0.33\columnwidth]{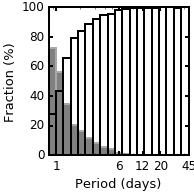}\\
  \includegraphics[width=0.33\columnwidth]{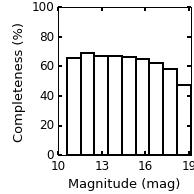}\includegraphics[width=0.33\columnwidth]{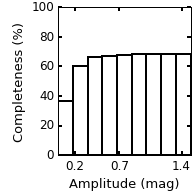}\includegraphics[width=0.33\columnwidth]{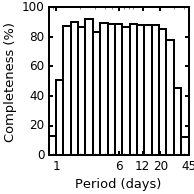}\\
  \caption{\label{fig:completeness}Period detection completeness and contamination levels for a mix of periodic
  synthetic light curves, with periods between 0.2 and 45 days, and amplitudes between 0.015 and 1.5 magnitudes.
Upper left panel shows $P_\mathrm{in}\times P_\mathrm{out}$ diagram, with every synthetic light curve shown as
a gray dot, recovered sample's periods shown in green, and selected sample's periods shown in blue. Harmonic
paths are shown as blue dotted line, and 1 day aliasing (cf. Equation \ref{eq:alias}) paths are shown as dashed
black lines. Bottom plots: Completeness distributions per magnitude (left), amplitude (medium) and period
(right) bins. Top: Given the selected sample, contamination levels are shown per amplitude (middle), and
period (right) bins. Fraction of true recovers are shown as empty bars, and fraction of contamination are
shown as filled gray bars.}
\end{figure}

\begin{figure}[bt]
  \centering
  \includegraphics[width=0.33\columnwidth]{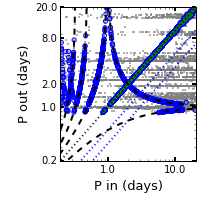} \includegraphics[width=0.33\columnwidth]{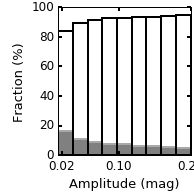}\includegraphics[width=0.33\columnwidth]{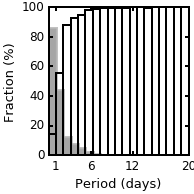}
  \includegraphics[width=0.33\columnwidth]{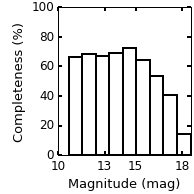}\includegraphics[width=0.33\columnwidth]{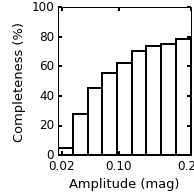}\includegraphics[width=0.33\columnwidth]{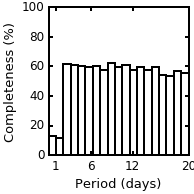}\\
\caption{\label{fig:completeness1} Same as Figure \ref{fig:completeness}, but for a mix of periodic synthetic 
light curves, with periods between 0.2 and \textbf{20 days}, and amplitudes between 0.015 and 0.2 magnitudes.}
\end{figure}

\textbf{We applied the same analysis and filtering to the synthetic light curves as was done for the
candidate member catalog.} Both input and output periods were recorded. From the results, two
samples were defined: synthetic light curves flagged as periodic were considered as in the
``selected''-sample; synthetic light curves were considered as in the ``recovered''-sample if the
input and output periods were the same, \emph{i.e.}, if the deviation between them was smaller than
10$\%$ the input period ($(|P_\mathrm{in}-P_\mathrm{out}|)\le 0.1P_\mathrm{in})$). The results for J
band in the two samples are shown in the $P_\mathrm{in}\times P_\mathrm{out}$ distribution in
Figures \ref{fig:completeness} and \ref{fig:completeness1}.

For the completeness analysis, a period in a synthetic light curve was considered successfully recovered if 
it was present in both selected and recovered samples. The completeness, \emph{i.e.,} successfully recovered 
periods divided by the number of input periods, for the first series of simulations is presented in Figure 
\ref{fig:completeness} (middle, top and bottom plots), where one can see that the completeness decreases 
slightly for increasing magnitude (plot at the bottom left), going from $\sim$69$\%$ for magnitudes around 
12, to 47$\%$ for magnitudes larger than 18. The completeness of the sample is quite uniform, and around 
$\sim$67$\%$ for amplitudes between 0.18 and 1.5 mag, having a significant drop between 0.015 and 0.18 
magnitudes, reaching only 37$\%$. It is also quite uniform, and around $\sim$87$\%$, for periods between 
1.2 and 24 days (bottom plot); but it drops to $\sim$12$\%$ for periods smaller than $\sim$1 day, and to 
$\sim$12$\%$ for periods larger than 36 days, which can be explained by strong aliasing factors in the 
very large, and very small period extremes.

Figure \ref{fig:completeness} also shows the contamination analysis for the selected-sample. In the top 
middle and right panels, 100$\%$ means the complete selected-sample, empty bars show the percentage of 
light-curves in this sample with successfully recovered periods per period or amplitude bins, and filled 
bars show the percentage of light-curves in the selected-sample, but that do not have
$(|P_\mathrm{in}-P_\mathrm{out}|)\le 0.1P_\mathrm{in})$ and, therefore, are contaminants to the sample. 
We advise the reader to keep in mind what are the samples used to build each histogram in Figure 
\ref{fig:completeness} and \ref{fig:completeness1}: while the completeness analysis takes into account the 
total number of periods available to be measured (\emph{i.e.}, every gray dot inside the 
$P_{in} \times P_{out}$); the contamination analysis accounts only for periods which were measured by the 
process described in the previous section (\emph{i.e.} blue and gray points inside the $P_{in} \times P_{out}$).
From the amplitude vs. contamination fraction histogram one can see that the contamination level is quite 
small for given amplitude, going from a maximum of about 16$\%$ for the smallest amplitude bin, and reaching 
only 5$\%$ for the highest amplitude bin. The period vs. contamination fraction histogram shows that the 
contamination level is insignificant ($\sim$1$\%$) for periods longer than 6 days, and that it increases 
significantly up to 70$\%$ for measured periods shorter than $\sim$1 day.

Even though the simulations presented in Figure \ref{fig:completeness} comprise the whole amplitude,and period
ranges measured for our periodic sample, they assume uniform distributions for the ranges considered, which is
unrealistic if we consider the real distributions of periods and amplitudes. Only $\sim$1$\%$ of the periodic
sample had measured periods longer than 20 days, and Figure \ref{fig:amplitude} shows that the measured
amplitudes are mostly low amplitudes up to 0.2 magnitudes, which corresponds to the first bin of amplitude
in Figure \ref{fig:completeness}. Therefore, we ran a second series of simulations, in order to consider
in more details a sample dominated by periods shorter than 20 days, and low amplitudes. 
 
The second series of simulations are presented in Figure \ref{fig:completeness1}. The completeness vs. 
amplitude panel shows that the sample's completeness increases from merely 4$\%$ for amplitudes smaller 
than 0.035 (which is very close to the data's 2$\%$ error level), to $\sim$78$\%$ for amplitudes larger 
than 0.18 mag in J filter. The completeness for given period is quite homogeneous, and about 57$\%$ for 
periods larger than 2 days, and as small as about 12$\%$ for periods smaller than 2 days. The completeness 
for given J magnitude bin is quite flat, and about 67$\%$ until magnitude 15, but it decreases toward 
larger magnitudes up to $\sim$14$\%$ for magnitudes larger than 18. The completeness increase with amplitude, 
and decrease with magnitude can be explained by the error distributions in Figure \ref{fig:error}, 
since higher amplitudes result in higher signal to noise ratio, while stars with larger magnitudes have 
larger errors, and therefore smaller signal to noise ratio.

\textbf{There is a small contamination level as a function of amplitude, which decreases from
  $\sim$16$\%$ to $\sim$5$\%$ with increasing amplitude.}
\textbf{From the upper left plot in Figure \ref{fig:completeness1}, one can see in the
  $P_\mathrm{out}>P_\mathrm{in}$ region that there exists some
  contamination arising from short $P_\mathrm{in}$ values aliased towards larger periods. The effect of this contamination can also be seen in 
  the region filled in gray inside the contamination fraction as a function of period plot (upper right): The contamination is very
  high for period bins up to $\sim$2 days, reaching $\sim$44$\%$ for periods between 1.0 and 2.0 days, and $\sim$86$\%$ for periods smaller than 1 
day. But it decrease to very small values for periods between 2 and 7 days, and it is almost
insignificant for periods larger than that. Additional sources of contamination will be discussed in
Appendix \ref{app:contamination}.
}

\subsection{Comparison With Periods in the Literature and Binary Stars Contamination}

\begin{figure}[bt]
  \centering
  \includegraphics[width=0.75\columnwidth]{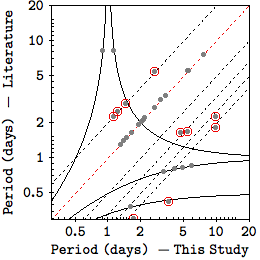}
  \caption{\label{fig:Plit}Comparison between periods measured in the present work, and in \citet{hend2011+}.
  Gray circles show the periodic stars in common between the two works. Stars with the same period measured
in both works fall over the red dashed line. Red circles show the eclipsing binaries candidates from
\citet{hend2011+}. Continuous black lines show 1 day aliasing paths, and
dashed black lines shows the path for harmonics.}
\end{figure}

\citet{hend2011+} observed CygOB2 in two seasons of 19 and 18 nights, with a few observations per 
night. They presented 121 stars variable in R and I band, 95 of which had measured periods. Within a 
distance of $0\farcs 45$ we found 79 of their variable stars also present in our catalogue: 7 of 
them were classified by the present study as non-variable stars; 9 were eclipse-like variables; 17 were 
classified as non-periodic variable stars; and 46 were periodic candidates. Figure \ref{fig:Plit} shows 
a comparison between the periods measures by \citet{hend2011+}, and the ones measured in the present study. 
Both studies measured periods of 32 stars in common, and the same period for 14 of them. Since we are only 
interested in periodic behaviour that may be reflecting the star's rotation, all the variables presented
by \citet{hend2011+} as eclipsing binary candidates (10 stars) were excluded from our periodic stars list. 
Taking out the eclipsing binary candidates from the plot in Figure \ref{fig:Plit}, all remaining stars 
without equal periods measured in both studies fall on the aliasing paths, and the 8 objects in this situation 
were also ruled out from our periodic list.

\subsection{Visual Light Curve Crosscheck}
\label{sec:visual}
Finally, we visually crosschecked the folded light curves for periodic stars selected in Section 
\ref{sec:periodic}, and verified if a periodic signal was indeed present on them. We considered as 
contaminants the stars that were selected as periodic but that did not show a visible periodic signal 
in the folded light curve. We found 23 (1.8$\%$ of the periodic sample )
contaminant stars this way. We also 
visually checked the light curves for the stars selected as periodic, but classified during de visual 
inspection as non-variable, and verified that 4 (0.3$\%$) of them did not show a periodic signal in the 
folded light curve, being considered as extra contaminations to the periodic sample.

\section{Results}
\label{sec:sec4}
\subsection{General Period Distributions}
\begin{figure}[tb]
  \centering
  \includegraphics[width=0.4\textwidth]{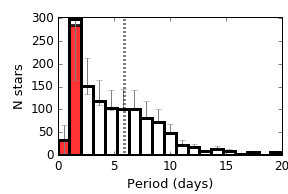}\\
  \includegraphics[width=0.4\textwidth]{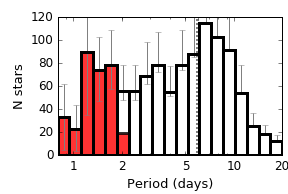}
  \caption{\label{fig:distP}Period distributions for the 1196 periodic stars with $P<20$ days found in the
    present study in both linear (top) and logarithm (bottom) scale. The median value for $P\geq2$ days stars 
    is shown as dotted line. \textbf{There were 879 stars with $P\geq2$ days and $P<20$ days. The red bins show the fast
rotators with $P<2$ days excluded from the analysis.}}
\end{figure}

After excluding eclipse-like variables, and possible visually-selected contaminants (see Appendix \ref{sec:visual}), we compiled a final 
list of 1211 candidate members of CygOB2 with periodic variability and periods between 0.86 and 32.49 days.
The period distribution for the 1196 stars with $P<20$ days is shown in Figure \ref{fig:distP}. Since only 
15 stars ($\sim1\%$ of periodic sample) have periods longer than 20 days, we kept those stars outside the 
distributions shown in the rest of the paper. Error bars were estimated taking into account the completeness 
and contamination analysis for periods up to 20 days (Figure \ref{fig:completeness1}): upper error bars take
into account the sample's incompleteness for each magnitude bin (Figure \ref{fig:completeness1}), and lower 
error bars take into account the contamination level per magnitude. Even though they were not used in the 
present results analysis, periods shorter than 2 days are shown in red in the distribution in Figure 
\ref{fig:distP}, and also in the following Figures in this Section. While not accounting for the shortest 
periods, the mean, standard deviation and median values for the general period distribution are 
$P$($\mu,\sigma, \nu$)= 6.67, 4.18, and 5.92 days.

\begin{figure}[tb]
  \centering
  \includegraphics[width=0.95\columnwidth]{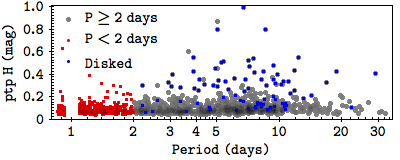}\\
  \includegraphics[width=0.95\columnwidth]{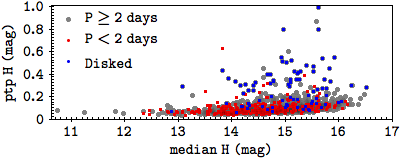}
  \caption{\label{fig:amp_P}Top: Period vs. ptp H amplitude for periodic stars. Bottom: median H magnitude 
  vs. ptp amplitude H for periodic stars. Stars with $P<2$ days are shown as red dots; Stars with $P\geq2$ 
days are shown as gray dots; Disked periodic stars are shown as blue dots.}
\end{figure}

The photometric ptp amplitudes for the H-band are shown as a function of periods in the top panel of 
Figure \ref{fig:amp_P}. The amplitude of 90$\%$ of the periodic stars is widely distributed between 0.03 
and 0.18 magnitudes; $9\%$ of them exhibit amplitudes between 0.18 and 0.51 magnitudes, and $1\%$ exhibit 
amplitudes as high as 1 magnitude. 

The bottom panel in Figure \ref{fig:amp_P} shows the photometric ptp amplitudes for the H-band as a function 
of magnitude. The increase of the lower amplitude envelope with magnitude is due to an observational bias 
related to the dependence of the minimum photometric errors on the magnitude (cf. Figure \ref{fig:error}),
which makes it increasingly difficult to detect low amplitude variables among fainter stars. On the other 
hand, the upper envelope of the distribution also shows an increase with amplitude, which can be explained 
by a larger and/or more uneven spot coverage for fainter (less massive) stars. There are 61 periodic stars 
outliers in the median H magnitude vs. amplitude distribution, showing higher amplitude than most of the 
stars with the same brightness. While 46$\%$ of the disked stars follow the main distribution, 74$\%$ of 
all the outliers are disked stars. This is consistent with the idea that most of periodicity observed in 
the light curves for disk-bearing stars arises mainly form two physical mechanisms: the rotational 
modulation by hot spots, which are expected to show higher variability amplitudes than cold spots 
\citep[e.g.][]{2001Carpenter,2009Scholz}, and by circumstellar obscuration, which may be the responsible
for the higher amplitudes observed \citep[e.g., AA Tauri stars;][]{2003Bouvier}. 

\begin{sidewaystable*}
  \caption{\label{tab:result}Rotational periods for CygOB2 periodic candidate members. Each column is
    explained in the text. A complete version of the Table containing the 894 periodic stars is available 
  electronically at CDS Strasbourg.}
 \tiny{
      \begin{tabular}{|p{2.1cm}p{0.8cm}p{0.8cm}|p{1.1cm}p{1.1cm}|p{0.8cm}p{0.8cm}|p{0.8cm}p{0.8cm}|p{0.8cm}p{0.8cm}
        p{0.7cm}|p{0.7cm}p{0.7cm}p{0.7cm}|p{0.7cm}p{0.7cm}p{0.7cm}|p{0.6cm}p{0.5cm}|}
%
    \hline
    Id & GDW13 &  GDW15 & RA & Dec & Stet& Per& Mass& Av &mJ &eJ& ptpJ &mH &eH &ptpH& mK& eK& ptpK& 
    Class\tablefootmark{a}& Disk\tablefootmark{b} \\
       &       &        & (h:m:s) & (d:m:s) & & (day) & ($\msun$) & (mag) & (mag) & (mag) & (mag) &
    (mag) & (mag) & (mag) & (mag) & (mag) & (mag) & & \\
    \hline
    \hline
  CygOB2-0000004 & 1047 &  & 20:32:04 & +41:14:19 & 0.46 & 8.35 & 0.49 & 4.29 & 16.61 & 0.08 & 0.03 &
    15.35 & 0.02 & 0.06 & 14.78 & 0.02 & 0.06 &  & \\
  CygOB2-0000007 & 1065 &  & 20:32:05 & +41:17:27 & 0.52 & 2.82 & 0.86 & 5.13 & 15.31 & 0.06 & 0.02 &
    14.23 & 0.02 & 0.05 & 13.74 & 0.02 & 0.06 &  & \\
  CygOB2-0000008 & 1064 &  & 20:32:05 & +41:12:48 & 0.79 & 3.05 & 0.79 & 2.74 & 15.66 & 0.08 & 0.02 &
    14.8 & 0.02 & 0.08 & 14.46 & 0.02 & 0.07 &  & \\
  CygOB2-0000010 & 1070 &  & 20:32:05 & +41:15:49 & 0.45 & 2.62 & 0.63 & 5.09 & 15.55 & 0.06 & 0.02 & 
    14.34 & 0.02 & 0.05 & 13.79 & 0.02 & 0.05 &  & \\
  CygOB2-0000015 & 1086 &  & 20:32:06 & +41:06:49 & 0.92 & 10.55 & 0.66 & 5.91 & 17.06 & 0.15 & 0.04
    & 15.72 & 0.03 & 0.13 & 15.04 & 0.03 & 0.12 &  & \\
  CygOB2-0000022 & 1113 &  & 20:32:07 & +41:16:01 & 1.11 & 2.18 & 0.6 & 5.17 & 15.02 & 0.1 & 0.02 & 
    13.83 & 0.02 & 0.08 & 13.29 & 0.02 & 0.08 &  & \\
  CygOB2-0000035 & 1159 &  & 20:32:10 & +41:07:45 & 0.61 & 7.55 & 0.31 & 4.26 & 16.59 & 0.09 & 0.03
    & 15.28 & 0.02 & 0.08 & 14.62 & 0.02 & 0.08 &  & \\
  CygOB2-0000037 & 5482 &  & 20:32:11 & +41:18:24 & 1.24 & 7.94 & 0.4 & 4.26 & 17.02 & 0.17 & 0.04
    & 15.77 & 0.03 & 0.14 & 15.18 & 0.03 & 0.13 &  & \\
  CygOB2-0000040 & 5483 &  & 20:32:11 & +41:07:52 & 0.83 & 7.81 & 0.36 & 4.05 & 17.47 & 0.15 & 0.05 
    & 16.18 & 0.03 & 0.13 & 15.56 & 0.03 & 0.12 &  & \\
  CygOB2-0000041 & 1175 & 70438 & 20:32:11 & +41:11:33 & 2.05 & 7.1 & 0.31 & 4.88 & 17.03 & 0.31 &
    0.04 & 15.71 & 0.03 & 0.28 & 14.88 & 0.02 & 0.22 & CL2 & 1\\
  CygOB2-0000044 & 1187 &  & 20:32:12 & +41:18:24 & 0.90 & 4.9 & 0.69 & 4.39 & 15.99 & 0.1 & 0.03 &
    14.84 & 0.02 & 0.08 & 14.32 & 0.02 & 0.08 &  & \\
  CygOB2-0000049 & 1208 &  & 20:32:12 & +41:16:47 & 0.75 & 7.44 & 0.53 & 4.24 & 17.15 & 0.13 & 0.04
    & 15.93 & 0.03 & 0.12 & 15.37 & 0.03 & 0.12 &  & \\
  CygOB2-0000052 & 1217 &  & 20:32:13 & +41:11:45 & 0.53 & 19.52 & 0.52 & 4.84 & 15.75 & 0.06 & 0.02
    & 14.46 & 0.02 & 0.06 & 13.87 & 0.02 & 0.05 &  & \\
  CygOB2-0000060 & 1229 &  & 20:32:14 & +41:16:47 & 0.32 & 5.17 & 0.66 & 3.88 & 15.59 & 0.06 & 0.02
    & 14.5 & 0.02 & 0.04 & 14.02 & 0.02 & 0.04 &  & \\
  CygOB2-0000066 & 1283 &  & 20:32:16 & +41:15:37 & 1.27 & 5.13 & 0.7 & 4.13 & 15.99 & 0.12 & 0.03
    & 14.85 & 0.02 & 0.11 & 14.35 & 0.02 & 0.09 &  & \\
  CygOB2-0000067 & 1287 &  & 20:32:16 & +41:18:28 & 1.31 & 9.55 & 0.63 & 3.67 & 16.38 & 0.13 & 0.03
    & 15.22 & 0.02 & 0.12 & 14.71 & 0.02 & 0.1 &  & \\
  CygOB2-0000070 & 5267 &  & 20:32:16 & +41:07:29 & 2.11 & 5.81 & 0.31 & 4.52 & 17.1 & 0.25 & 0.04
    & 15.78 & 0.03 & 0.24 & 15.06 & 0.03 & 0.24 &  & \\
  CygOB2-0000072 & 1299 &  & 20:32:16 & +41:14:39 & 1.78 & 7.62 & 0.55 & 5.0 & 16.07 & 0.18 & 0.03 
    & 14.75 & 0.02 & 0.17 & 14.06 & 0.02 & 0.16 &  & \\
  CygOB2-0000074 & 1310 &  & 20:32:17 & +41:18:49 & 0.81 & 6.57 & 0.41 & 3.12 & 17.17 & 0.13 & 0.04
    & 16.01 & 0.03 & 0.12 & 15.49 & 0.03 & 0.11 &  & \\
  CygOB2-0000078 & 5485 &  & 20:32:17 & +41:15:03 & 0.55 & 7.16 & 0.26 & 3.99 & 16.92 & 0.12 & 0.04
    & 15.7 & 0.03 & 0.09 & 15.07 & 0.03 & 0.08 &  & \\
  CygOB2-0000079 & 1328 &  & 20:32:17 & +41:17:58 & 2.09 & 2.31 & 0.58 & 4.1 & 15.32 & 0.17 & 0.02 
    & 14.16 & 0.02 & 0.16 & 13.63 & 0.02 & 0.15 &  & \\
  CygOB2-0000083 & 1341 &  & 20:32:18 & +41:14:04 & 0.89 & 5.85 & 0.42 & 4.35 & 16.51 & 0.11 & 0.03
    & 15.24 & 0.02 & 0.09 & 14.64 & 0.02 & 0.08 &  & \\
  CygOB2-0000086 & 5269 &  & 20:32:19 & +41:05:55 & 1.09 & 6.13 & 0.29 & 4.2 & 17.23 & 0.16 & 0.04
    & 15.89 & 0.03 & 0.14 & 15.23 & 0.03 & 0.13 &  & \\
  CygOB2-0000089 & 1368 &  & 20:32:20 & +41:12:00 & 0.38 & 4.2 & 0.64 & 3.74 & 15.43 & 0.06 & 0.02
    & 14.4 & 0.02 & 0.04 & 13.91 & 0.02 & 0.04 &  & \\
  CygOB2-0000090 & 5487 &  & 20:32:20 & +41:13:33 & 1.00 & 3.79 &  &  & 17.43 & 0.17 & 0.05 & 16.1
    & 0.03 & 0.17 & 15.44 & 0.03 & 0.14 &  & \\
  CygOB2-0000092 & 5488 &  & 20:32:20 & +41:13:31 & 0.79 & 4.45 & 0.17 & 3.66 & 17.35 & 0.17 & 0.05 
    & 16.0 & 0.03 & 0.13 & 15.34 & 0.03 & 0.11 &  & \\
  CygOB2-0000093 & 1384 &  & 20:32:20 & +41:16:54 & 0.51 & 8.88 & 0.31 & 3.52 & 16.53 & 0.07 & 0.03
    & 15.27 & 0.02 & 0.07 & 14.69 & 0.02 & 0.06 &  & \\
  CygOB2-0000102 & 1427 &  & 20:32:22 & +41:14:47 & 0.70 & 9.02 & 0.52 & 4.09 & 15.76 & 0.08 & 0.02
    & 14.48 & 0.02 & 0.06 & 13.92 & 0.02 & 0.06 &  & \\
  CygOB2-0000121 & 1498 &  & 20:32:25 & +41:17:45 & 1.01 & 4.23 & 0.58 & 2.99 & 15.46 & 0.1 & 0.02
    & 14.37 & 0.02 & 0.08 & 13.88 & 0.02 & 0.07 &  & \\
  CygOB2-0000126 & 1516 &  & 20:32:26 & +41:16:44 & 0.45 & 6.16 & 0.61 & 3.31 & 15.61 & 0.05 & 0.02
    & 14.5 & 0.02 & 0.04 & 14.0 & 0.02 & 0.05 &  & \\
  CygOB2-0000131 & 1520 &  & 20:32:27 & +41:07:17 & 1.18 & 4.63 & 0.47 & 5.5 & 15.74 & 0.11 & 0.02
    & 14.28 & 0.02 & 0.09 & 13.53 & 0.02 & 0.07 &  & \\
  CygOB2-0000134 & 1532 &  & 20:32:27 & +41:12:01 & 1.05 & 5.32 & 0.75 & 6.21 & 16.0 & 0.12 & 0.03 
    & 14.73 & 0.02 & 0.1 & 14.1 & 0.02 & 0.08 &  & \\
  CygOB2-0000142 & 1555 &  & 20:32:28 & +41:14:39 & 0.56 & 4.09 & 0.68 & 7.16 & 16.45 & 0.09 & 0.03
    & 14.85 & 0.02 & 0.07 & 13.98 & 0.02 & 0.06 &  & \\
  CygOB2-0000143 & 5280 &  & 20:32:28 & +41:09:00 & 0.80 & 4.96 & 0.46 & 6.12 & 16.85 & 0.14 & 0.04
    & 15.39 & 0.02 & 0.11 & 14.65 & 0.02 & 0.09 &  & \\
  CygOB2-0000156 & 1589 &  & 20:32:30 & +41:16:51 & 0.48 & 2.17 & 0.57 & 4.91 & 15.4 & 0.05 & 0.02
    & 14.23 & 0.02 & 0.05 & 13.67 & 0.02 & 0.05 &  & \\
  CygOB2-0000163 & 1613 &  & 20:32:30 & +41:17:45 & 1.13 & 8.05 & 0.8 & 4.12 & 15.82 & 0.12 & 0.02 
    & 14.74 & 0.02 & 0.1 & 14.26 & 0.02 & 0.08 &  & \\
  CygOB2-0000165 & 1615 &  & 20:32:31 & +41:12:19 & 0.94 & 8.27 & 0.63 & 5.93 & 16.76 & 0.13 & 0.03 
    & 15.4 & 0.02 & 0.11 & 14.76 & 0.02 & 0.09 &  & \\
  CygOB2-0000176 & 1663 &  & 20:32:33 & +41:09:59 & 1.00 & 8.34 & 0.44 & 5.1 & 16.69 & 0.14 & 0.03 
    & 15.33 & 0.02 & 0.11 & 14.66 & 0.02 & 0.09 &  & \\
  CygOB2-0000178 & 1679 &  & 20:32:33 & +41:11:33 & 0.92 & 6.69 & 0.69 & 6.84 & 16.85 & 0.14 & 0.04 
    & 15.4 & 0.02 & 0.11 & 14.7 & 0.02 & 0.09 &  & \\
  CygOB2-0000183 & 1694 &  & 20:32:34 & +41:17:19 & 1.00 & 4.64 & 0.39 & 3.76 & 16.35 & 0.12 & 0.03 
    & 15.17 & 0.02 & 0.11 & 14.6 & 0.02 & 0.09 &  & \\
    \hline
    \hline
  \end{tabular}
  \tablefoottext{a}{Disk IR evolutionary status from GDW13 attributed by using \citet{Wilking2001+} scheme. 
    CL1: Class 1 YSO. CL2: Class 2 YSO. FS: Flat Spectrum. PTD: pre transitional disks. Ha: H$\alpha$ 
    emitter according to GDW13 or to \citet{2008Vink}. BWE: blue stars with excesses candidate stars with 
    disks, but with optical color bluer than the cluster locus). lowmass :	low-mass disk with excesses 
    only in [8.0] and [24]. high-incl:	highly inclined disk with excesses only in [8.0] and [24]
}
\tablefoottext{b}{Disk presence: 1 if true.}
}
\end{sidewaystable*}

The 894 periodic stars with $P>2$ days are listed in Table \ref{tab:result} which shows their ID inside our
variability survey catalogue, their ID in the catalogues presented by GDW13, and GDW15, their coordinates in
the present study; Stetson variability Index (Section \ref{sec:sec3}), Period (Section \ref{sec:periodic});
Mass, and Av (Section \ref{sec:mass}), median J,H, and K magnitudes, their propagated errors, and their ptp
amplitude for each band (Section \ref{sec:sec3}); their IR-class according to GDW13 (Section \ref{sec:sec2}
, and \ref{sec:res:disk}), and a flag indicating the presence of disk, assuming the value 1 for disked
stars, and 0 for non disked stars. 

\subsection{Period Distribution for Disk-Bearing vs. Non-Disked Stars}
\label{sec:res:disk}
To further investigate the nature of the period distribution presented in Figure \ref{fig:distP}, we
compared the period distributions of disk-bearing and non-disked stars. Since we do not have reliable
mass accretion rate measurements for Cygnus OB2 low mass stars\footnote{GDW13 used IPHAS data in order 
  to select H$\alpha$ emitters inside the r'-H$\alpha$ vs. r'-i' colour-colour diagram. They identified
  52 sources in such diagram, but only one of them was selected as a periodic star.}, we rely on X-ray 
  emission and IR disk diagnosis in order to distinguish between CTTs and WTTs samples, and
investigate the star/disk connection effect in the period distributions. The disked stars in the sample
are the disk-bearing stars listed by GDW13. The Non-Disked stars in the sample are the X-ray emitter
sources from \citet{Wright2014b+} classified as members by \citet{2016Kashyap} without IR-excess.

\begin{figure*}[tb]
  \centering
  \includegraphics[width=0.4\textwidth]{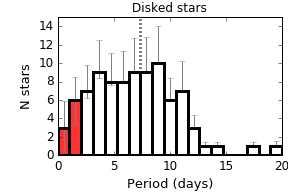}
  \includegraphics[width=0.4\textwidth]{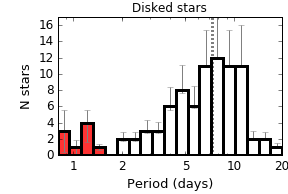}\\
  \includegraphics[width=0.4\textwidth]{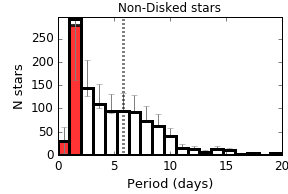}
  \includegraphics[width=0.4\textwidth]{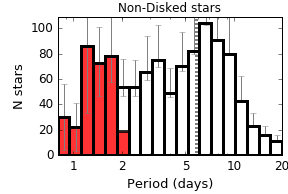}\\
  \caption{\label{fig:Pdisk}Period distributions for periodic disk-bearing stars with $P\leq20$ days
    (top) and periodic non-disked stars with $P\leq20$ days (bottom) for both linear and logarithm
    distributions. The median value for $P\geq2$ days stars is shown as dotted line.\textbf{ There were 80
    disked stars, and 799 non-disked stars with with $P\geq2$ days and $P<20$ days. The red bins
show the fast rotators with $P<2$ days excluded from the analysis}.}
\end{figure*}

There were 91 disk-bearing periodic stars, 82 of them with $P\geq2$ days, and 1120 non-disked periodic 
stars, 812 of them with $P\geq2$\footnote{Note that these numbers are slightly different
from the numbers presented in Figure \ref{fig:Pdisk}. This is because the plot in the refereed
Figure presents only stars with $P\leq20$ days, and there are 3 disked and 13 non-disked stars with
periods larger than that.}. The period distributions for disk-bearing and non-disked
stars are shown in Figure \ref{fig:Pdisk}. The period distributions for disk-bearing and non-disked stars are 
shown in Figure \ref{fig:Pdisk}. A visual inspection of the distributions suggests that the two samples
exhibit different rotational properties, even without taking into account the bins with $P<2$ days. The 
disk-bearing stars distribution is quite flat for periods in the range 4-11 days, decreasing for
periods smaller than 4 days. The non-disked stars distribution is flat for periods between 4 and 7 days, 
but it decreases for periods larger than that, with a larger fraction of stars with periods smaller than 
4 days.

The mean, standard deviation and median values are $P$($\mu,\sigma, \nu$)=  7.87, 4.36, and 7.34 
days for disk-bearing stars, and $P$($\mu,\sigma, \nu$)= 6.55, 4.14, and 5.80 days for non-disked 
stars. A visual inspection in the distributions of Figure \ref{fig:Pdisk}, and a quick look to the 
statistical values suggest that although both distributions show a large dispersion, disked stars 
are on average rotating slower than non-disked stars. A Kolmogorov-Smirnov (KS) test\footnote{
ks$\_$2samp from Python package scipy.stats} considering only periods higher than 2 days statistically 
supports this idea, with a probability of only 0.4$\%$ that the two distributions came from the same 
parent distribution. 

\subsection{Period Distributions for Given Masses}
\subsubsection{Mass Estimation}
\label{sec:mass}
In order to estimate masses for the candidate members it would be necessary the knowledge of
individual extinctions, which in turn would require the knowledge of spectral types for the stars. 
Since there is no available information about the spectral type for the low mass members coming 
from spectroscopic studies in the literature, individual extinction for candidate members could 
not be formally determined. An alternative for that is to look for optical counterparts in other 
surveys, and use optical colours to estimate the reddening of each object. Since colours are not 
affected by the distance, a distance independent individual $A_V$ can be estimated for each star 
by using colour-colour diagrams (CCD). This can be done by using a chosen extinction law in order 
to estimate the displacement of a star in the CCD from an appropriate isochrone with zero extinction, 
plotted in the same diagram. 

\begin{figure}[bt]
  \centering
  \includegraphics[width=0.4\textwidth]{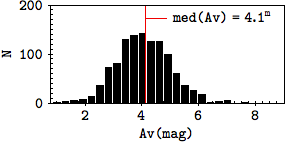}\\
  \includegraphics[width=0.45\textwidth]{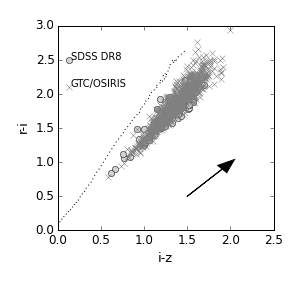}
  \caption{\label{fig:Av}Top: Individual $A_V$ distribution for periodic stars. The
    red line shows the median value, 4.1$^m$. Bottom: riz CCD for periodic stars. \citet{Covey} 
    empirical dwarf sequence is show as dotted lines. A black arrow shows a reddening vector 
    from \citet{1998Schlegel} for $A_V$=2$^m$.
  }
\end{figure}

As described in Section \ref{sec:literature}, \emph{riz} photometry is available from GDW12, who give
\emph{riz} indices for 1086 stars of our periodic sample, with error smaller than 0.2 magnitudes in 
each filter: 991 from GTC/OSIRIS observations, and 95 from SDSS DR8. Thus, we used the i-z vs. r-i
colour-colour diagram in order to estimate individual extinctions. Instead of using a 3.5 Myr 
\citet{2000Siess} isochrone to estimate $A_V$, as in GDW12, we used an empirical dwarf sequence 
from \citet{Covey}. We adopted relative extinction for riz bands from \citet{1998Schlegel}\footnote{
  Transformed from u'g'r'i'z' (USNO 40 in) to ugriz (SDSS 2.5m) according to 
  \url{http://classic.sdss.org/dr7/algorithms/jeg_photometric_eq_dr1.html}
}, where the authors used a $R_V$=3.1 extinction law from \citet{1989Cardelli+} to evaluate relative 
extinctions. Since both \citet{Covey} dwarf sequence, and the riz data used from GDW12 (cf. Section 
5.1 of their paper) are in SDSS photometric system, no transformation was required, and this justifies 
the differences in our method for estimating individual extinctions, and the one used by GDW12.
The CCD for \emph{riz} colours is shown on the bottom plot of Figure \ref{fig:Av}.

We were able to estimate $A_V$ for 1058 periodic stars, 971 using GTC/OSIRIS data, and 86 using SDSS 
DR8 data. The distribution of individual extinctions obtained is shown in the top of Figure 
\ref{fig:Av}. The median $A_V$ obtained for the periodic sample was 4.1$^m$, which is in good 
accordance with the 4.33$^m$ value obtained by GDW12 for stars in CygOB2 center.

\begin{figure}[bt]
  \centering
  \includegraphics[width=0.4\textwidth]{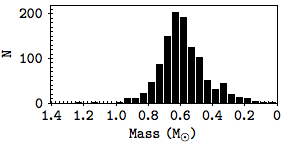}\\
  \includegraphics[width=0.45\textwidth]{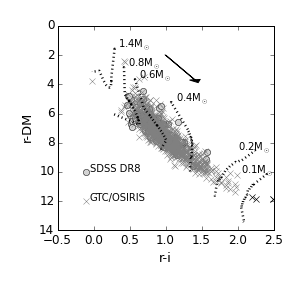}
  \caption{\label{fig:massa}Top: histogram with mass distribution for periodic stars. Bottom 
    r-i vs. r-DM CMD for periodic stars. Mass tracks from \citet{2014Bell} for Pisa models 
    \citep{2012Tognelli} with semi-empirical corrections are show as dotted lines. A black arrow 
    shows a reddening vector from \citet{1998Schlegel} for $A_V$=2$^m$. Black Xs show 
    stars excluded from the mass estimation because they were too far away from
    the minimum mass track.
  }
\end{figure}

We then used our estimations of individual extinction values and the r-i vs. r CMD to estimate
masses. We adopted a distance modulus DM=10.62 \citep[d=1.33kpc][]{2015Kiminki}. For each periodic
star with estimated $A_V$, we de-redden it for the individual $A_V$, and applied the distance
modulus. Masses were then estimated given the position of the de-redden, and distance-corrected star
inside the CMD in relation to a grid of PMS semi-empirical isochrones from \citet{2014Bell}, with
PISA models \citep{2011Tognelli,2012Tognelli} for solar metallicity ($Z_\odot$=0.013), and ages 
in the range 0.1-30 Myr.
The grid of semi-empirical isochrones was built by using the CMDfit software\footnote{CMDfit software,
  author: Tim Naylor: \url{http://www.astro.ex.ac.uk/people/timn/tau-squared/software.html}}, and the
  bolometric corrections applied were calculated by the software's authors by folding spectra with 
  opacities from BT-Settl \citep{2011Allard} through the desired filter response and applying empirical 
  corrections from \citet{2014Bell}. In our case, we chose the filter responses for SDSS filters
  \citep{2010Doi} with an AB zero point. The isochrones used in the grid comprise masses in the 
  range 0.1-8$\msun$. We estimate masses for 1054 periodic stars.
  The bottom plot in Figure 
  \ref{fig:massa} shows a r-i vs.
  r CMD for de-redden periodic stars and mass tracks from 0.1 to 1.4$\msun$. The middle plot in Figure 
  \ref{fig:massa} shows a histogram with the distribution of estimated masses for periodic stars. 
  The CMD in Figure \ref{fig:massa} also shows a lack of stars with masses larger than $\sim$0.9$\msun$, 
  which is mainly due to the fact that most of the stars with masses larger than that are very close to the 
  saturation limit in our WFCAM/UKIRT observations.

A caveat on using optical colors for estimating masses is that stars with disks may be affected by accretion,
scattering, or obscuration of the central star by the inner disk. Only 65 stars in our sample are periodic, 
have disks and had masses estimated as described in this section. We verified that only 3 of those stars had 
evolutionary status from GDW13 compatible with some ongoing phenomena that could affect their optical colours: 
1 star has H$\alpha$ emission, 1 star has colours compatible with scattering, and 1 star has a disk in high 
inclination. Since they correspond to a very small fraction of our sample, we consider that these effects 
do not influence our mass estimation.

\subsubsection{Period Distributions in Different Mass Ranges}
\label{sec:res:mass}
\begin{figure}[tb]
  \centering
  \includegraphics[width=0.45\textwidth]{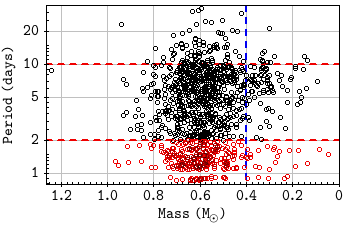}
  \caption{\label{fig:P_M_ALL} Mass vs. period distribution for periodic candidate members of CygOB2.
    Stars with reliable periods ($P\geq2$ days) are shown as black circles, and stars with dubious periods 
  ($P<2$ days) are shown as red circles. Dashed red lines delimit for periods equal to 2 and 10 days.
The blue dashed line marks the limit between the mass bins analysed in the present section.}
\end{figure}

\begin{figure}[bt]
  \centering
  \includegraphics[width=0.4\textwidth]{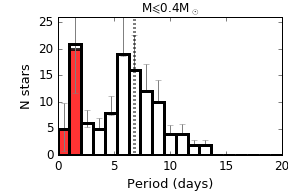}
  \includegraphics[width=0.4\textwidth]{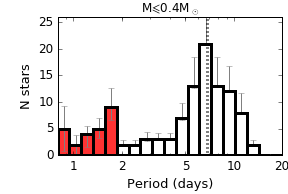}
  \includegraphics[width=0.4\textwidth]{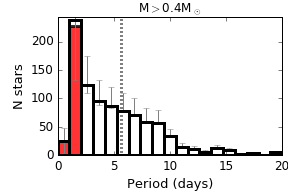}
  \includegraphics[width=0.4\textwidth]{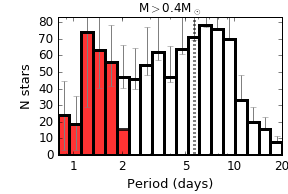}
  \caption{\label{fig:P_Mrange} Histograms showing period distributions for given mass bins in
    linear (left) and logarithm scale (right). $M\leq0.4M\msun$ (top), and $M>0.4\msun$ (bottom).
    The median value for $P\geq2$ days stars is shown as dotted line. textbf{There were 89 stars with
    $M\leq0.4M\msun$ and $P\geq2$ days and $P<20$ days, and 676 stars with $M>0.4\msun$ in the same
period interval. The red bins show the fast rotators with $P<2$ days excluded from the analysis.}}
\end{figure}

Mass estimations for PMS stars can be extremely model dependent, but for all models, lower effective
temperatures correspond to lower mass stars, and thus the sense of variation of rotation with mass
is model independent \citep{2001Herbst}. To test a possible mass-rotation connection in our data, we
split the periodic sample into two mass bins: $M\leq0.4\msun$ (90 stars with $P\geq2.0$ days), and
$M>0.4\msun$ (687 stars with $P\geq2.0$ days).  Histograms showing the period distribution for each
of the three mass bins are shown in Figure \ref{fig:P_Mrange}. From the distributions, it is evident
that the period distribution for medium, and slower rotators present a mass dependence. The period
mean, standard deviation and medium values are: $P$($\mu,\sigma, \nu$)= 7.0, 2.9, and 6.9 days for
the $M\leq0.4\msun$ sample, and $P$($\mu,\sigma, \nu$)=  6.5, 4.2, 5.7 days for the $M>0.4\msun$
sample.

Figure \ref{fig:P_M_ALL} shows the mass vs. rotational period distribution for Cygnus OB2. The blue dashed
line marks the limits between the mass bins adopted, and are meant to allow a simple comparison between Figures
\ref{fig:P_M_ALL} and \ref{fig:P_Mrange}. Keeping in mind that our data sampling does not allow us
to access the complete fast rotators population, a lack of periods in the range 2-5 days can be verified 
for the $M\leq0.4\msun$ mass bin. Both linear and log scaled rotational period distributions are strongly
peaked for the lower mass interval, but the peak, around 6.1 days, is more evident in log-scaled distribution. 
The overall distribution gets broader for the $M>0.4\msun$ mass bin, which is reflected as an increase in 
the  distribution's $\sigma$. The peaked distribution verified for lower masses is less evident here, and 
an increase in the number of faster and intermediate rotators makes the period distribution flatter.

A KS-test gives a probability of $\sim0.02\%$ that the distributions for $M\leq0.4\msun$ and 
$M>0.4\msun$ samples came from the same parent population. The results therefore indicate that
for the periodic sample analysed here, the lower mass stars are rotating on average slower 
than higher mass stars. 

Since masses derived from CMD are highly model dependent, we verified that our results hold for 
masses estimated from different models. We repeated the process  for estimating masses described 
in Section \ref{sec:mass} using 3 other models: \citet{1998Baraffe}, and \citet{2015Baraffe} with 
\citet{2014Bell} empirical corrections, and \citet{2000Siess}. The first two models result 
in masses very similar to the ones we adopted. For \citet{2000Siess} model, masses larger than
0.4$\msun$, PISA models yield masses $\sim1.2$ times larger. For masses bigger than 
0.4$\msun$ PISA and \citet{2000Siess} can be different up to a factor of 2. 

\textbf{We adopted the mass limit of 0.4 $\msun$ because of the following reason: The convective
  boundary, \emph{i.e.}, the mass under which all the stars will be fully convective even in the MS,
  is around $\sim$0.3 $\msun$ for the model used. On the other hand, all low mass PMS stars are
  initially fully convective. \citet{2012Gregory} estimated the age at which a PMS star first
  develops a radiative core for the Pisa models, and according to their estimates a 0.5 $\msun$ star
  develops its radiative core around $\sim$9.3 Myr, and a 0.4 $\msun$ star around $\sim$15 Myr. As
  we discuss in Section \ref{sec:lit}, CygOB low mass stars age estimates are between 2.5 - 6.75 Myr
  \citep{wright2010+}, but these estimates were based on \citet{2000Siess} models, which, when
applied to PMS stars, may result in underestimated ages down to a 1.5-2.0 factor. Accounting for
this, we split the sample at 0.4 $\msun$ as a way to guarantee that the lower mass sample will not
be contaminated by stars that are no longer fully convective.}

\textbf{We also investigated the effect of other choices of mass limit and concluded that the results
  hold for different values. Using 0.5 $\msun$ gives results qualitatively similar to 0.4 $\msun$ 
  (lower mass stars rotate slower) and a KS-test yield
  statistically different samples. When splitting the periodic stars in several mass ranges, the
  overall result is kept the same: Stars with $M<$0.4 $\msun$ rotate slower than stars in the mass
  range 0.4 - 0.6 $\msun$ and a KS-test results in 0.05$\%$ chance they came from the same parent
  distribution. The 0.4 - 0.6 $\msun$ and 0.6 - 0.8 $\msun$ samples have very similar rotational
  properties and can not be distinguished according to a KS-test. When comparing 0.4 - 0.5 $\msun$ 
  sample with 0.5 - 0.8 $\msun$ sample, the former is slower than the later, but according to the 
  KS-test it is not possible to say if they are different, which supports our choice of 0.4 $\msun$
  as mass limit for the comparison.
}

\section{Discussion}
\label{sec:sec5}

Given the CygOB2 rich population and the fraction of candidate members with periodic variability, 
Cygnus OB2 is a valuable target for testing the theory of stellar angular momentum evolution 
during the PMS. 

In this Section we discuss our results in the context of early rotational evolution for low mass 
stars. It is of utmost importance to keep in mind the limitations of our sample. Since the 
faster rotators ($P<2$days) in the sample are strongly contaminated, our discussion is based 
on a few fast rotators, intermediate and slow rotators. As in \citet{2002Herbst}, we based our
definition of rotational regimes on values of rotational angular velocity. Given that rotational
angular velocity relates to the measured periods as $\omega=\frac{2\pi}{P}$, we called stars with
$\omega\leq0.5\mathrm{\frac{rad}{day}}$ ($P>12.56$ days) very slow rotators, 
0.5$\mathrm{\frac{rad}{day}}<\omega\leq1.0\mathrm{\frac{rad}{day}}$ ($P>6.28$days) slow rotators,
and stars with $\omega>2\mathrm{\frac{rad}{day}}$ (P$<$3.14days) fast rotators. Median rotators 
are stars rotating with rotational angular velocity in the range 
1$\mathrm{\frac{rad}{day}}\leq\omega\leq2\mathrm{\frac{rad}{day}}$.

\subsection{Does CygOB2 Corroborate the Disk-Locking Scenario?}

In Section \ref{sec:res:disk} we considered as CTTS the stars listed by GDW13 as disk-bearing stars,
and as WTTS, stars selected as candidate members by having X-ray emission from \citet{2016Kashyap},
and not listed as disk-bearing stars. A caveat arises from this selection procedure, since it does
not guarantee that the disk-bearing stars are still interacting with their disks, nor account for
disked stars with inclinations that do not produce IR-excess. Nonetheless, using this criterion the
disk-fraction in the full candidate member sample is 24$\%$\footnote{As reference, the disk-fraction
of similarly aged CepOB3b is 33$\%$ \citep{2012Allen}}, and about 10$\%$ among candidate periodic
stars.  Given that the sample of stars with reliable period measured (with P$\geq$2 days) contains
only 82 stars with disks, the disk fraction in this sample is only 7.5$\%$. When looking at the
light-curve morphological classification for the non-disked sample: 39.1$\%$ of the stars were
classified as periodic candidates, 1.1$\%$ as eclipse-like, 14.5$\%$ as non-periodic variable stars,
and 45.3$\%$ as non-variable stars. For the disk-bearing stars: 13.8$\%$ were periodic candidates,
5.7$\%$ were eclipse-like, 59.7$\%$ were non-periodic variables, and 10.8$\%$ where non-variables.
This indicates that data-sampling used here is more efficient for detecting periodicity among
non-disked stars, or equivalently, that the sample is biased towards WTTS.

The origin of the bias towards WTTS can be explained by taking into account the different physical 
mechanisms responsible for variability in CTTS and WTTS. Within WTTS we expect to detect mainly stars 
with variability caused by cold spots, which are expected to produce a low amplitude JHK variability 
(typically smaller than one tenth of magnitude), even for large spot coverages 
\citep[e.g.][]{2001Carpenter}. For the CTTS the variability scenario may be more complicated. Besides 
the variability caused by the presence of cold spots, the most common sources of variability in CTTS 
are: obscuration by circumstellar material, accretion driven variability (like the presence of hot 
spots and variable mass accretion rates), and instabilities in the accretion disk 
\citep[e.g.][]{2001Carpenter,2014Cody,2014Stauffer,2015Rice,2015McGinnis,2016Sousa,2017Roquette}.
While sometimes a single physical process may dominate the star's light curve, the existence of several
concurrent variability sources is often the case. Adding up common limitation in the datasets, as 
limited time and wavelength coverage, multiple physical process composing a complex light curve may
not be distinguishable. Thus, rotational periods in CTTS are often masked by other variability sources
in the light curve. 

This bias towards WTTS has been reported by other studies in the literature 
\citep[e.g.,][]{2004Cohen,2005Herbst}, and it is assumed to be present in all the studies comparing the
rotation of CTTS and WTTS. In the present study, a direct consequence of it is that the size of the CTTS
sample is much smaller than the size of the WTTS sample. Consequently the results 
regarding disked stars are less statistically significant. For this reason, we did not compare CTTS and 
WTTS separating them by mass ranges. Therefore, we could not verify the statistics correlation of 
rotational periods with disk-diagnosis for restricted mass ranges, in order to test the evidences that
disk-locking acts differently in different mass ranges. 

Another bias arising from the use of disk diagnosis based on IR-excess is that a correlation of rotation 
with disk presence diagnosed via IR-excess can be a secondary effect due to dependence of IR-excess on 
mass suggested by some authors \citep{Littlefair2005}. \citet{1998Hillenbrand} showed that the IR-excess 
produced by the disk is a function of the disk properties, but also of the star's mass and radius. 
The contrast between the disk and the star's photosphere is larger for higher mass stars, so with lower 
contrast it is more difficult to detect disks of lower mass stars. In regions where lower mass stars rotate
faster than higher mass stars \citep[e.g. NGC 2362][]{Irwin2008+}, lower mass stars with undetected disks
can mimic a correlation between fast rotators and non-disked stars. This effect can be minimized by using
longer IR wavelengths to identify stars with IR-excess, and \citet{2007CiezaBaliber} showed that since 
the photosphere/disk contrast is higher in the mid-IR, disks can be detected even for lower mass stars 
by using Spitzer/IRAC colours. We discarded the possibility of our sample being affected by such bias.
The analysis of rotation as a function of mass bin presented in Section \ref{sec:res:mass} showed that,
contrary to several other young regions studied, lower mass stars in CygOB2 rotate on average slower 
than higher mass stars. Given that even if disks around some lower mass stars were undetectable within
the limits of the Spitzer data used by GDW13 for evaluating disk-bearing stars, the contamination by 
those stars would introduce slow rotators to the non-disked sample. This would produce a contamination 
in the sense of occluding the correlation between IR-excess and slow rotators. Hence we rule out the possibility 
that a correlation between IR-excess and slow rotators in our sample could be mimicked by a secondary
effect.
 
Focusing on the whole list of members with periodic variability in CygOB2 (see Section \ref{sec:res:disk}), 
and even though our dataset is only complete for rotators with $P\geq2$days, our results corroborate the
general idea that the star-disk interaction has some influence in the rotational regulation of young 
stars. Disked stars in our sample are rotating on average slower than non-disked stars. 
 
\begin{figure}[tb]
  \centering
  \includegraphics[width=0.4\textwidth]{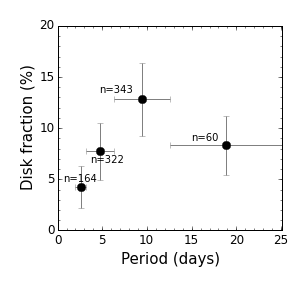}\\
  \includegraphics[width=0.4\textwidth]{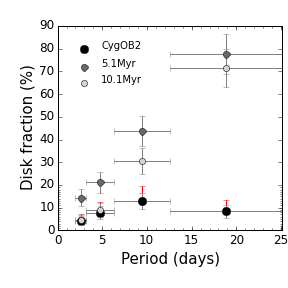}\\
  \caption{\label{fig:P_diskfrac} Disk fraction as a function of period. In both plots, circles show 
  the center of each period bin, while horizontal bars indicate the size of the bin. The bins show: 
fast, intermediate, slow, and very slow rotators. Observational data from the present study are shown 
as black. Top: The number of stars in each bin is show for each point. Bottom: disk fractions resulting
from M2 simulations from \citet{2015VasconcelosBouvier} for 5.1 Myr, and 10.1 Myr are shown together with 
the fraction observed in CygOB2.}
\end{figure}

\citet{2007CiezaBaliber} present an alternative way of looking for observational evidences for the 
disk-locking phenomenon by studying the disk-fraction as a function of period. A plot of disk-fraction
per period bin is shown in Figure \ref{fig:P_diskfrac} (left). The black circle shows the central 
period for each bin, and the bars show the length of each bin. The bins were built considering fast 
rotators, intermediate rotators, slow rotators, and very slow rotators defined as in the beginning of 
the present Section. Even though the disk fraction for periodic stars is low ($\sim9\%$), the top 
plot of Figure \ref{fig:P_diskfrac} shows that the disk fraction is actually 
quite dependent on the rotational period, having a maximum value of $\sim13\%$ for stars with rotational
periods between 8.4 days, and 12.6 days.
 
\citet{2015VasconcelosBouvier} investigated the disk fraction variations as a function of period for 
simulated populations under disk-locking effect with ages between 1.0 Myr and 12.1 Myr. Figure 7 of their 
paper shows, for a model starting with different period distributions for disked and non-disked stars, 
that the disk fraction increases with increasing rotational periods at all ages. Their M2 model has, at
age 5.1 Myr, a disk fraction varying from 15$\%$ to 77$\%$ for periods from 2 days to 18 days, and from 
5$\%$ to 72$\%$ at 10.1 Myr. Their results for 5.1 Myr, and 10.1 Myr are shown in the bottom plot in 
Figure \ref{fig:P_diskfrac}, together with the results for our sample. In both plots in Figure 
\ref{fig:P_diskfrac} vertical bars follow standard errors of a Poisson counting, as the error bars used 
by \citet{2015VasconcelosBouvier}. 
Results from \citet[][hereafter GDW16]{gdw16} investigation on the disk-survival in CygOB2 suggest that
the environmental
feedback on disk evolution for CygOB2 members may be responsible for a decrease of about 20$\%$ on the 
disk fraction. The red vertical error bars in the bottom plot in Figure \ref{fig:P_diskfrac} shows a tentative 
correction for this effect and adds up 20$\%$ of each point's disk fraction. Remembering that the disked
periodic sample is biased due the variability mechanisms acting on disked stars, as described early in 
the present section, our dataset seems to corroborate the disk-locking results for periods up to 12.6 days,
even though the increase of the disk fraction with period occurs in a less accentuate way when compared 
with \citet{2015VasconcelosBouvier} results. For periods longer than that, \emph{i.e.} for the slowest 
rotators bin in the plots of Figure \ref{fig:P_diskfrac}, the disk-fraction decreases, what may 
indicate that our dataset suffers from contamination from field stars. 

A possible explanation for the lower disk-fraction compared to the models could be a strong premature 
disk loss due to the high energy radiation coming from the OB stars in the association, and Figure 
\ref{fig:P_diskfrac} (top panel) would be  showing signatures of a primordial 
disk-locking with a reduced disk-fraction due to fast disk dissipation in the association. In this case,
the disk lifetime distribution in CygOB2 would be very different from the ones used by 
\citet{2015VasconcelosBouvier}, since they assume in their simulations that most stars are born with 
disks, and suffer from a smooth decrease on disk-fraction with time. Consequently, a comparison with
their results would not be possible. Other possible explanations would be that CygOB2 low mass population 
is a few Myr older than previous thought, or that the environmental conditions in CygOB2 make the disk
fraction decrease with time more steeply than considered in \citet{2015VasconcelosBouvier}. It seem 
reasonable that combination of the two effects could explain the differences between CygOB2 data, 
and simulated data for the fast, slow, and intermediate rotators points in the right Figure \ref{fig:P_diskfrac}.

Since Figure \ref{fig:P_diskfrac} suggests that the very slow rotator sample suffers from strong 
contamination, we re-examine the results in Section \ref{sec:res:disk} by removing stars with 
$P>12.56$ days from the analysis. By doing so, we found that $P$($\mu,\sigma, \nu$)=  7.0, 2.7,
and 7.1 days for disked stars, and  $P$($\mu,\sigma, \nu$)= 5.7, 2.6, and 5.5 days for non-disked
stars. A KS-test gives an 0.0001$\%$ change that the two distributions came from the same parent
distribution, showing that the results from Section \ref{sec:res:disk} hold even when excluding
the very slow rotators from the analysis.
\subsection{CygOB2 Inside the Picture of PMS Rotational Evolution}
\label{sec:lit}

\begin{figure}[tb]
  \centering
  \includegraphics[width=0.45\textwidth]{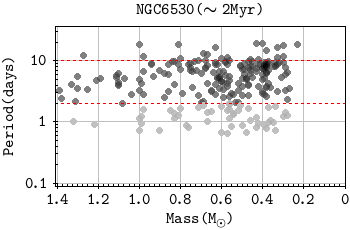}
  \includegraphics[width=0.45\textwidth]{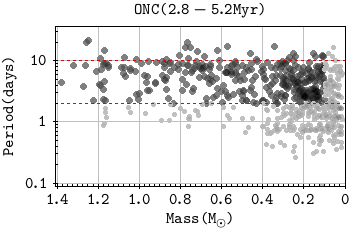}
  \includegraphics[width=0.45\textwidth]{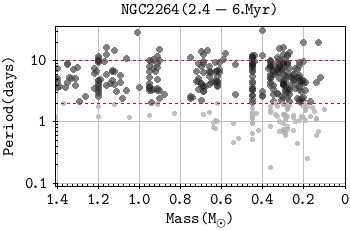}
  \includegraphics[width=0.45\textwidth]{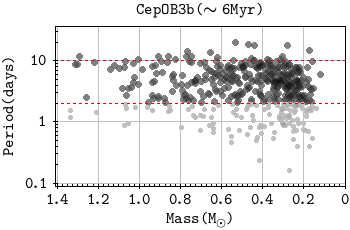}
  \includegraphics[width=0.45\textwidth]{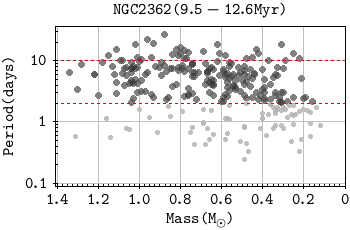}
  \includegraphics[width=0.45\textwidth]{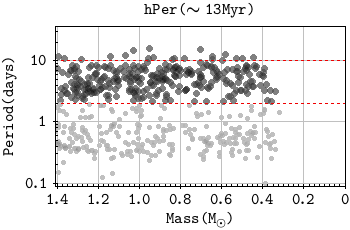}
\caption{\label{fig:P_M_lit} Rotational periods as a function of mass for several young regions 
  with ages between 1-15 Myr. Dashed red lines delimit the periods between 2 and 10 days. Stars 
used for comparison with the present study are shown as dark gray.}
\end{figure}

Figure \ref{fig:P_M_lit} presents distributions of mass vs. rotational period for several 
young regions. These regions were selected from the list presented by 
\citet{BouvierMatt2013ReviewII} in their review on the evolution of the AM in young low mass stars. 
The clusters and associations chosen have ages up to 15 Myr and rotational period
samples that are numerous enough to be considered statistically significant. The selected regions
are: NGC 6530 \citep{2012HendersonStassun}, with 244 measured periods in the mass range 0.2-2.0
$\msun$; Orion Nebulae Cluster \citep[ONC,][]{IrwinBouvier2009,RodriguezLedesma2009}, with 528 
measured periods in the mass 
range 0.015-1.4$\msun$; NGC 2264 \citep{2005Lamm,Affer2013,2016Venuti}, with about 581 measured 
periods in the mass range 0.2-3.0$\msun$; CepOB3b \citep{Littlefair2010+} in the mass range 
0.1-1.3$\msun$, with 460 measured periods; NGC 2362 \citep{Irwin2008+}, with 271 measured periods
in the mass range 0.1-1.2$\msun$; and hPer \citep{2013Moraux}, with 586 measured periods in the 
mass range 0.4-1.4$\msun$. 

The panels in Figure \ref{fig:P_M_lit} are presented in order of age. For each cluster, we 
adopted the most recent age estimation available in the literature: NGC 6530 has $\sim$2 Myr
\citep{2013Bell}, ONC has 2.8-5.2 Myr \citep{2009Naylor}, NGC 2264 has 2.4-6 Myr \citep{2009Naylor},
CepOB3b has $\sim$6 Myr \citep{2013Bell}, NGC2362 has 9.5-12.6 Myr \citep{2013Bell}, and hPer has
13 Myr \citep{2008MayneNaylor}. Those age estimates partially explain the different rotational
scenarios observed in CepOB3b, and NGC 2362 by \citet{Littlefair2010+}, since the reviewed ages
put the two regions in different evolutionary stages. 

From the mass vs. period plots in Figure \ref{fig:P_M_lit} one can see a large spread in period 
for all ages. The youngest region is NGC6530, and it presents rotational periods widely scattered 
between 0.5 days and 19 days. The oldest one is hPer, and its distributions, also widely scattered,
presents a lower envelope at periods around 0.3 days, and maximum rotational periods around 15 days.
While in NGC 6530 only 32$\%$ of the sample is composed by fast rotators, for hPer the fast 
rotators percentage is $\sim$56$\%$. ONC, NGC 2264, CepOB3b, and NGC 2362 seem to present some
transitional properties from the rotational state of NGC6530 to the rotational state of hPer. For
some clusters (ONC, CepOB3b, and NGC 2362), the spin up of the distribution's lower envelope seems
to be more efficient for lower mass stars. 
  
\begin{figure}[tb]
  \centering
  \includegraphics[width=0.45\textwidth]{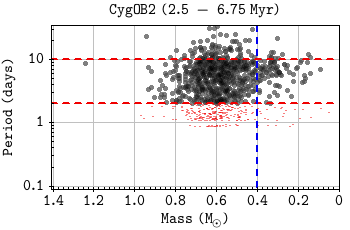}
  \caption{\label{fig:P_M_scale} Mass vs. period distribution for periodic candidate members of 
  CygOB2. Stars with reliable periods ($P\geq2$ days) are shown as black circles, and stars with 
dubious periods ($P<2$ days) are shown as red circles. Dashed red lines delimit the periods
between 2 and 10 days, while blue dashed lines delimits the mass bin analysed in the present section.}
\end{figure}

In order to make the comparison between the samples presented in Figure \ref{fig:P_M_lit}, and
CygOB2 sample (Figure \ref{fig:P_M_scale}), we filtered each sample for stars with masses larger
than 1.4M$_\odot$, and smaller than 0.1M$_\odot$, and periods smaller than 2 days. Stars filtered 
by this selection rule are shown as light gray circles in the mass vs. period plots in Figure
\ref{fig:P_M_lit}. Figure \ref{fig:P_M_scale} shows the same plot as in Figure \ref{fig:P_M_ALL},
but in the same scale as the plots in Figure \ref{fig:P_M_lit}. After filtering each sample, their 
sizes are: 187 stars 
in NGC6530 sample; 351 in ONC; 288 in NGC2264; 342 in CepOB3b; 198 in NGC2362; and 309 in hPer.
Against 894 stars in CygOB2.

\citet{wright2010+} derived ages between 2.75 and 6.75 Myr for CygOB2 low mass population, 
with a median value of 3.5 Myr for the center field, and 5 Myr for a northwestern field. 
For deriving stellar properties, they used \citet{2000Siess} models converted to 2MASS 
photometric system using \citet{1995KenyonHartmann}. Given that the methods used by 
\citet{2008MayneNaylor}, \citet{2009Naylor}, and \citet{2013Bell} for evaluating stellar 
parameters result in ages 1.5-2.0 greater than ages estimated by previous methods, we 
stress that comparisons with other clusters in the literature 
using CygOB2 age as a parameter must be done with caution. That being said, using the age range 
as the unique criterion would place CygOB2 somewhere between ONC, and NGC2362. 

One feature that markedly varies from cluster to cluster in the plots of Figures \ref{fig:P_M_lit},
and \ref{fig:P_M_scale}, is the slope of the distribution's upper envelope for lower mass stars.
\citet{Irwin2008+} suggest that this varying slope for masses between 0.1-0.5M$_\odot$ was a 
result of cluster evolution with age. Following this suggestion, \citet{2012HendersonStassun} 
used such slope as an age proxy, and suggest NGC6530 would be in an earlier evolutionary stage 
compared to ONC. They argued that this younger age would justify the remarkable differences
between the mass vs. period distributions for NGC6530 and other clusters. In special, they 
found that lower mass stars in NGC6530 are rotating much slower than the higher mass stars, 
unlike observed in most other young clusters where lower mass stars rotate faster than higher
mass stars. The authors argued that given an younger age, the explanation for the opposite 
sense in the period-mass relationship, would be that the lowest mass stars in the cluster 
are currently spinning up, and that in a future step of evolution they would end up rotating 
faster than the higher mass ones. 

We found in Section \ref{sec:res:mass} that lower mass stars in CygOB2 also rotate slower than 
higher mass stars. Given the much older age of CygOB2, compared to NGC 6530, we argued that the
different age of these clusters could not explain the observational results of lower mass stars 
rotating on average slower than higher mass stars. An alternative explanation to this inverse 
period-mass relation for lower mass stars would be the role of environment conditions on 
regulating the stellar AM.

\subsection{Are the Mass vs. Periods Relations Sensitive to the Environment?}

A mass-rotation connection was first reported for stars in ONC (2.5-5.2 Myr), where 
\citet{2001Herbst} found that while stars in the cluster with spectral types between M2 and K 
presented a bi-modal distribution with one peak around fast rotators ($\sim$2 days), and another 
around slow rotators ($\sim$8 days); stars with spectral type later than M2 presented a unimodal
distribution with a single peak around fast rotators ($\sim$2 days). Given the observed differences
in the period distribution, and given that the stars with later spectral type were on average 
faster rotators than the earlier, the authors called attention to a possible mass-rotation 
connection. Studies for other regions like NGC 2264 \citep{2005Lamm}, and IC 348 \citep{Littlefair2005}
favoured this connection.

A role of the environment in determining the mass-rotation relation has been suggested by
\citet{Littlefair2010+}. They found that even though lower mass stars rotate faster than
higher mass stars in CepOB3b, this difference is less strong than in other clusters, since 
CepOB3b has an excess of slow rotating low mass stars compared to other regions.

As is the case in NGC 6530 \citep{2012HendersonStassun}, lower mass stars in CygOB2 rotate on 
average slower than higher mass stars (Section \ref{sec:res:mass}) with a median rotational 
period of 6.9 days for lower mass stars, and 5.7 days for higher mass stars. Since our data is 
incomplete 
for fast rotators, we do not have enough information to discuss the existence of bi-modality in
the period distributions, but apart from that, the shape of period distributions for different
mass ranges presented on Figure \ref{fig:P_Mrange} shows clear sign of a mass dependence: while 
period distribution for higher mass stars (M$\geq$0.4M$_\odot$) is quite flat for periods
between 2-9 days, with a long tail towards slower rotators, it is peaked around slow rotators 
for the lower mass stars ($M\leq0.4\msun$) in the sample showing a lack of fast rotators.

Before putting these results in context, we examine the possibility that a bias due to underestimated
extinction could affect the results. An underestimated extinction would cause highly extincted higher
mass stars to be considered as low mass stars. Since differences in the period distributions for 
lower and higher mass stars have been repeatedly reported in the literature, including studies for 
regions with relatively low and homogeneous extinction - as is the case of NGC2264 
\citep[$A_V\sim0.4$][]{2008Dahm}, where this contamination effect is probably minimum, higher mass
stars mistaken by lower mass stars would bring contamination to the lower mass period sample. In
regions where higher mass stars are slower rotators than lower mass stars, this bias would introduce
slow rotators to the lower mass period distributions which was initially composed of faster rotators,
biasing the distribution towards slower rotators, and masking the differences between the two
mass-regime distributions. In the case of higher mass stars rotating on average faster than lower 
mass stars, this bias would introduce fast rotators to the lower mass stars distribution and the 
distribution would be biased towards such faster rotators. In both cases, we would expect the period 
distribution for lower mass stars to be somehow flattened due to the contamination. When comparing
CygOB2 Mass vs. Period distributions (Figure \ref{fig:P_M_scale}) with the other regions (Figure 
\ref{fig:P_M_lit}), one can see that CygOB2 presents a clear lack of fast and intermediate rotators 
within its lower mass stars, and therefore we ruled out the possibility that the result may be mimicked 
by a contamination due to higher mass stars with underestimated extinction. A third possibility that 
remains is that the extinction for the stars in our sample is underestimated as whole, and in this 
case our observations would not reach lower mass at all, and the whole lower mass bin would actually 
be composed by higher mass stars with underestimated masses. Since we saw in Section 
\ref{sec:res:mass} that a KS-test between the samples with $M\leq0.4\msun$, and $M\geq0.4\msun$
returned a probability of only $\sim$0.02$\%$ that the two samples came from the same parent 
population, we also excluded the hypothesis that the smaller mass sample could be a sub-sample 
of higher mass stars with underestimated extinction.

A final caveat on interpreting our results arises from the fact our sample is incomplete for fast
rotators with periods under 2 days. To check if the statistical difference between the rotational
scenario of different mass regimes holds after removing the faster rotators from the sample, we 
present in Table \ref{tab:median} the mean, standard deviation, and median values $(\mu,\sigma,\nu)$
for the period distributions of the young regions presented in Figure \ref{fig:P_M_lit} for both 
full, and P$\geq$2 days samples. We divided each cluster sample in two mass ranges using 
$M=0.4\msun$ as a cut. For each cluster, the mass regime that rotates slower on average is 
stressed in bold. Since hPer sample includes only a few stars with masses $M\leq0.4\msun$, 
we present the results from this cluster in Table \ref{tab:median}, but we did not use it for the 
present investigation. From Table \ref{tab:median}, one can see that filtering the samples for 
$P\geq2$ days does not change qualitatively the rotation-mass connection. Nevertheless, for NGC
2264, KS-tests between the samples of higher and lower mass stars without faster rotators than 
$P=2$ days, changes significantly. The last column of Table \ref{tab:median} shows a KS-test 
between the lower mass stars from each clusters, with low mass stars in CygOB2, and shows that 
the lower mass end of CygOB2 Mass vs. Period distribution is dissimilar to all the other clusters 
considered. 

\begin{table*}
    \caption{\label{tab:median} Period mean, standard deviation, and median values $(\mu,\sigma,\nu)$
      for each cluster presented in Figure \ref{fig:P_M_lit} for the full, and $P\geq2$ days samples.
  \tablefoottext{a}{p-value from KS-test between the samples $M<0.4\msun$, and $M\geq0.4\msun$, given 
  in percentage, meaning the probability that the two sample were derived from the same parent
distribution.}
  \tablefoottext{b}{Sample composed with only 25 stars.}
\tablefoottext{c}{Sample composed with only 9 stars.}}
 \tiny{
   \centering{
   \begin{tabular}{r|p{1.7cm}p{1.7cm}|p{1.7cm}p{1.7cm}|p{1.5cm}p{1.5cm}|p{2.5cm}}
     & \multicolumn{2}{c|}{$(\mu,\sigma,\nu)$}  & \multicolumn{2}{c|}{$(\mu,\sigma,\nu)$} & \multicolumn{3}{c}{KS-test ($\%$)\tablefootmark{a}} \\
     \hline
     & \multicolumn{2}{c|}{M$<0.4$M$_\odot$} &  \multicolumn{2}{c|}{M$\geq$0.4M$_\odot$} &\multicolumn{2}{c}{between mass ranges}& (with CygOB2 \\
     & \multicolumn{2}{c|}{(days)} &  \multicolumn{2}{c|}{(days)} & &   &for M$<0.4$M$_\odot$) \\
     \hline
     Region & All & P$\geq$2days & All & P$\geq$2days & All & P$\geq$2days  & P$\geq$2days \\
    \hline
    CygOB2 & - & \textbf{(7.0,2.9,6.8)} & - & (6.5,4.2,5.7) & - &  0.02 & - \\
    \hline
    \hline
    NGC6530  & \textbf{(6.5,4.2,6.8)}   & \textbf{(6.4,3.3,5.6)}  & (5.3,3.7,4.7) & (6.5, 3.4, 5.6) & 1.4 & 2.1 & 0.16 \\
    ONC & (3.3,2.8,2.5)  & (5.0, 2.7, 4.4)  & \textbf{(5.7,3.9,5.4)}  & \textbf{(6.6, 3.6, 6.4)} & 10$^{-9}$ & 0.0001 & 10$^{-9}$\\
    NGC2264 & (4.6,3.7,3.9) &  (6.2, 3.4, 5.2)& \textbf{(5.4,4.1,4.3)}  & \textbf{(6.3, 4.0 5.1)} & 0.26 & 74.26 & 0.02 \\
    CepOB3b & (4.0,2.7,3.5) & (5.1, 2.5, 4.6) & \textbf{(4.9,3.5,4.4)}  & \textbf{(6.0, 3.2, 5.3)} & 1.16 & 1.83 & 10$^{-7}$\\
    NGC2362 & (3.0,3.2,1.9) & (5.0, 3.8, 3.0) & \textbf{(5.8,4.4,4.9)}  & \textbf{(6.9, 4.1 ,6.1)} & 10$^{-8}$ & 0.072 & 0.002\\
    hPer & (2.0,2.0,1.4)\tablefootmark{c}  & (4.2, 1.8, 3.3)\tablefootmark{d}  & \textbf{(3.3,3.0,2.7)}  & \textbf{(5.5, 2.5, 5.1)} & - & - & -\\
\hline
  \end{tabular}
}
}
\end{table*}

The observational results for CygOB2 could be explained if the lower mass stars in the sample were 
remaining locked to their disks for a longer time than the higher mass stars in the sample (solar-
type). One 
possible reason why the lower mass star could keep their disks for a longer time is if primordial 
mass segregation occurs, \emph{i.e.}, very low mass stars are more widely distributed than solar-type 
stars. They would thus lie further away from the ionizing radiation of central OB stars than the more
concentrated solar-type stars. They would as well be in lower stellar density regions, thus avoiding 
disk-disruptive encounters. However, \citet{Wright2014b+}, and \citet{2016WrightDance} found no 
signal of mass segregation in the association, and additionally GDW16 showed that close encounters 
in CygOB2 is not important in regulating disk dissipation, so this explanation may not apply. 

The present study shows that, like NGC 6530, CygOB2 presents qualitatively a mass-rotation connection
in the opposite sense from other clusters, with lower mass stars rotating on average slower than
higher mass stars. We consider that the statistical differences between the distributions from the
two regions are due to different rotational ages, since NGC 6530 is about 2 Myr old, and CygOB2 stars
have ages ranging from 2.5 to 6.75 Myr. Also due to this age difference between the two regions, we
refute the explanation given by \citet{2012HendersonStassun} based on youth to explain lower mass
stars being slower rotators than higher mass stars in the region, since for CygOB2 this 
explanation would not apply. Instead, we raise the hypothesis that the environmental influence 
on regulating the rotation may be a better explanation. 

Since both CepOB3b, and CygOB2 are OB associations with similar ages, it would be reasonable to
expect they would have similar rotational properties. Even though they both present an excess of
slow rotators within their lower mass members, their period distributions are statistically different,
and their rotational-mass connection is qualitatively inverse. Those differences could be explained 
by a smaller concentration of O stars: CygOB2 is a notoriously massive OB association, with more than
160 confirmed OB stars \citep{2015WrightMassive} within its members, 73 of which are O stars, while 
CepOB3b is a small association with a massive population composed by only 3 O stars, and 33 B stars 
\citep{1959Blaauw,1964Blaauw} stars spread over $\sim$10 pc \citep{1964Blaauw}. On the other hand,
NGC 6530 is a core cluster of the Sgr OB1 association and it is located in the eastern part of the
very bright Lagoon Nebula \citep{2000Sung}. NGC 6530 is also 3-4 times richer in OB stars than ONC 
\citep{2004Damiani}, whose population has similar age. 

\subsubsection{Does CygOB2 Massive Population Regulate Low Mass Star Rotation?}

OB stars can influence their environment due to their strong UV field. Far ultraviolet (FUV) photons 
(6eV$<h\nu<$13.6eV) can dissociate H$_2$ molecules, and extreme ultraviolet (EUV) photons 
($h\nu>$13.6eV) are capable of ionizing hydrogen atoms. Because of that, regions with intense local 
UV field can be hostile to the evolution of circumstellar disks, and to the processes of star formation 
\citep[e.g.][]{1998Johnstone,2004Adams,2010Guarcello,gdw16}. In special, GDW16 recently found evidences 
that disks are more rapidly dissipated in regions of CygOB2 with intense local UV.

To test the effect of CygOB2 massive stars on the rotational properties of nearby YSO we investigated 
how the rotational period distributions vary as a function of local UV field. To estimate local UV fluxes,
we used the technique adopted in GDW16 and \citet{2007Guarcello}: We propagated the FUV and EUV fluxes 
emitted by each O star to the position of each periodic stars using 2D projected distances. The effect 
of using 2D projected distances instead of real distances is discussed on section 3.1 of GDW16, and it 
was shown by the authors to have very small impact in the analysis.

 \begin{table*}[tb]
   \caption{\label{tab:UVdisk}For each UV flux sample and for samples of disk-bearing and non-disked 
     stars: Mean, standard deviation and median for each period distribution. Number of stars in the 
     sample (N), KS-test between disk-bearing and non disked stars. Samples selected given FUV incident
     flux are shown in the top of the table, and samples selected given EUV incident flux are shown
   in the bottom table.}
   \centering
   \begin{tabular}{p{1.2cm}|p{1.8cm}|p{0.4cm}|p{1.8cm}|p{1.4cm}|p{1.0cm}}
     \cline{2-5}
     &  \multicolumn{2}{|c|}{Disk-bearing}  &  \multicolumn{2}{c|}{Non-disked} & \\
     \cline{2-6}
FUV     &  $(\mu,\sigma,\nu)$ & N   &   $(\mu,\sigma,\nu)$ & N   &   \multicolumn{1}{c|}{ KS-test}\\
     \cline{1-6}
     \multicolumn{1}{|c|}{   low UV}  &  8.2, 4.6, 7.6  &  43 & 6.5, 4.0, 5.6  & 343  &    \multicolumn{1}{c|}{ 0.1$\%$}    \\
     \multicolumn{1}{|c|}{   high UV} & 7.2, 3.6, 7.1 & 29 & 6.6, 4.3, 6.0 & 469  &    \multicolumn{1}{c|}{57$\%$}\\
     \hline
     \cline{1-6}
   \end{tabular}

   \vspace{1mm}

   \begin{tabular}{p{1.2cm}|p{1.8cm}|p{0.4cm}|p{1.8cm}|p{1.4cm}|p{1.0cm}}
     \cline{2-5}
     &  \multicolumn{2}{|c|}{Disk-bearing}  &  \multicolumn{2}{c|}{Non-disked} & \\
     \cline{2-6}
EUV     &  $(\mu,\sigma,\nu)$ & N   &   $(\mu,\sigma,\nu)$ & N   &    \multicolumn{1}{c|}{KS-test}\\
     \cline{1-6}
     \multicolumn{1}{|c|}{   low UV} &  8.2, 4.5, 7.6 & 56 &  6.4, 4.1, 5.5 &  406 &    \multicolumn{1}{c|}{0.01$\%$} \\
     \multicolumn{1}{|c|}{   high UV} & 7.1, 3.8, 6.2 & 26 & 6.7, 4.2, 6.2  & 406 &   \multicolumn{1}{c|}{ 81$\%$} \\
     \cline{1-6}
\end{tabular}
\end{table*}

The UV flux emitted by each of the 73 O stars, and 3 Wolf-Rayet (WR) stars in CygOB2 was estimated by GDW16 
(Table 1 in their study). Their estimates for FUV flux is presented in terms of Habing flux 
G$_0$=$1.6\times10^{-3}$erg/cm$^2$/s\footnote{For reference: The average UV flux in the spectra range 
912-2000\AA $\;$ in the solar neighborhood is 1.7G$_0$ \citep{1968Habing}}, and their EUV fluxes
\footnote{number of ionizing photons with $\lambda<912$\AA$\;$  per second per cm$^2$.} are in 
photons/s/cm$^2$.  A map for incident FUV and EUV fluxes for CygOB2 candidate member stars was 
presented by the authors in their Figure 3. Using their estimates for the O stars UV flux, we 
calculated the incident UV flux at the position of each periodic star. As in GDW16, B stars 
are omitted because their census is still incomplete and their contribution to the whole UV field 
in the association is negligible compared to the O and WR stars. 

\begin{figure*}[bt]
  \centering
{$\log(F_{FUV})\leq3.7G_0$ \hspace{3.0cm}   $\log(F_{EUV})\leq11.42$photons/s/cm$^2$ }\par\medskip  
\vspace{-0.2cm}
   \includegraphics[width=0.4\textwidth]{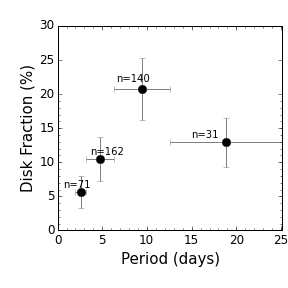}
   \includegraphics[width=0.4\textwidth]{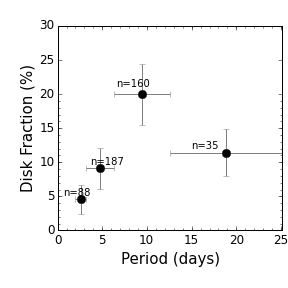}\\
  { $\log(F_{FUV})>3.7G_0$ \hspace{3.0cm}  $\log(F_{EUV})> 11.42$photons/s/cm$^2$}\par\medskip
   \includegraphics[width=0.4\textwidth]{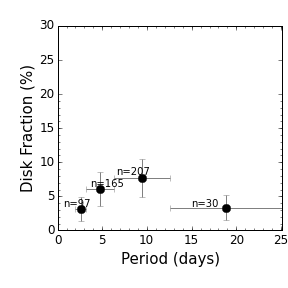}
   \includegraphics[width=0.4\textwidth]{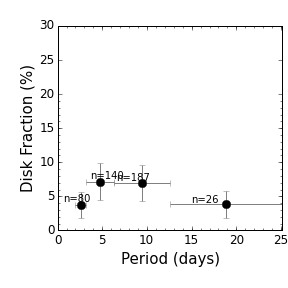}
   \caption{\label{fig:diskUV}Same as Figure \ref{fig:P_diskfrac} but for periodic stars with low UV 
     incident radiation (top). and stars with high UV incident radiation (bottom).}
 \end{figure*}

Using the FUV and EUV local fluxes, we define as regions with low UV incidence star positions 
where $\log(F_{FUV})\leq3.7G_0$ or $\log(F_{EUV})\leq 11.42$photons/s/cm$^2$, and regions with high
UV incidence star positions where $\log(F_{FUV})>3.7G_0$ or $\log(F_{EUV})> 11.42$photons/s/cm$^2$
\footnote{We also tested more extreme values as $\log(F_{FUV})=4.3G_0$, and 
  $\log(F_{EUV})=12.0$photons/s/cm$^2$ and verified that they produce the same results qualitatively. 
}.
. 
Disk fraction as a function of rotational period plots for stars in regions 
with high and low UV incidence are shown in Figure \ref{fig:diskUV}. From the plots, one can see that 
the maximum disk fraction goes from 21$\%$ for slow rotators stars in low FUV incidence regions, to 
7.3$\%$ for slow rotators in high FUV incidence regions; and from 20.3$\%$ for slow rotators in low 
EUV incidence regions, to 7.1$\%$ for intermediate rotators in high EUV incidence regions. Figure 
\ref{fig:diskUV} suggest that high incident UV does yield faster disk dissipation, even 
though it does not change qualitatively the trend of increasing disk fraction for longer periods. 

While GDW16 results suggest that regions with high UV incident fields can rapidly erode disks, our results 
suggest that this can directly influence disk-rotation connection. 
Table \ref{tab:UVdisk} shows the $(\mu,\sigma,\nu)$ values for each period sample for disk-bearing 
and non-disked stars in each UV incidence sample. The number of stars in each sample, and a KS-test between 
samples of disk-bearing and non-disked stars are also included in the Table. From the table one can see 
that while the disk-bearing and non-disked stars samples are different for regions with low UV incidence 
(a KS-test gives 0.1$\%$ probability they came from the same parent distribution for FUV samples, and 0.01$\%$ 
probability for EUV sample), in regions with high UV incidence a disk-rotation connection can not be verified 
and results from KS-test does not discard the possibility that disk-bearing and non-disked period distributions 
came from the same parent distribution in such samples.  

 \begin{table*}[tb]
   \caption{\label{tab:UV}For each UV flux sample and for each mass range considered: Mean, 
     standard deviation and median for each period distribution. Number of stars in the 
     sample (N), KS-test between mass-range. Samples selected given FUV incident flux are 
     shown in the top of the table, and samples selected given EUV incident flux are shown
   in the bottom table.}
   \centering
   \begin{tabular}{p{1.3cm}|p{1.8cm}|p{0.4cm}|p{1.8cm}|p{1.4cm}|p{1.0cm}}
     \cline{2-5}
     &  \multicolumn{2}{|c|}{$M\leq0.4M_\odot$}  &  \multicolumn{2}{c|}{$M>0.4M_\odot$} & \\
     \cline{2-6}
FUV     &  $(\mu,\sigma,\nu)$ & N   &   $(\mu,\sigma,\nu)$ & N   &   \multicolumn{1}{c|}{ KS-test}\\
     \cline{1-6}
     \multicolumn{1}{|c|}{   low UV}  &  6.9 2.6, 6.3  & 36  & 6.9, 4.3, 5.9  & 241  &    \multicolumn{1}{c|}{ 5$\%$}    \\
     \multicolumn{1}{|c|}{   high UV} &  7.3, 3.0, 7.1 & 50 & 6.5, 4.3, 5.7  &  388 &    \multicolumn{1}{c|}{0.2$\%$}\\
     \hline
     \cline{1-6}
\end{tabular}

\vspace{1mm}

\begin{tabular}{p{1.3cm}|p{1.8cm}|p{0.4cm}|p{1.8cm}|p{1.4cm}|p{1.0cm}}
     \cline{2-5}
    &  \multicolumn{2}{|c|}{$M\leq0.4M_\odot$}  &  \multicolumn{2}{c|}{$M>0.4M_\odot$} & \\
     \cline{2-6}
EUV     &  $(\mu,\sigma,\nu)$ & N   &   $(\mu,\sigma,\nu)$ & N   &    \multicolumn{1}{c|}{KS-test}\\
     \cline{1-6}
     \multicolumn{1}{|c|}{   low UV} &  7.2, 3.4, 6.5  & 37 &   6.7, 4.4, 5.8  & 297 &    \multicolumn{1}{c|}{1.1$\%$} \\
     \multicolumn{1}{|c|}{   high UV} & 7.1, 2.4, 7.1 & 49 &  6.6, 4.3, 5.7  & 332 &   \multicolumn{1}{c|}{ 0.5$\%$} \\
          \cline{1-6}
\end{tabular}
\end{table*}

Next, like in section \ref{sec:res:mass}, we built period distributions for different mass ranges for 
low and high UV incident radiation. The shape of the distribution do not seem to be significantly 
affected by the amount of UV incident radiation. Both low and high local UV incidence samples show 
a mass-rotation connection with lower mass  stars rotating in average slower than higher mass stars.
Table \ref{tab:UV} shows the $(\mu,\sigma,\nu)$ 
values for each distribution. The number of stars in each sample, and a KS-test between the two 
mass range for each sub-sample is also shown. We performed KS-tests between each pair of samples 
for the same mass range and regions with low and high UV incidence, but for all pairs of samples 
the KS-test yielded that we can not exclude the possibility that the two samples came from the same 
parent distribution. From Table \ref{tab:UV}, one can see that KS-tests between period distributions
for different mass ranges in samples with high incident UV flux give smaller probabilities that 
the two samples came from the same parent distribution than regions with low incident UV. This 
difference, even if less strong than the one found for the case of disk-rotation connection, may 
hint that UV incident fluxes can also influence the mass-rotation connection. 

\textbf{The magnitude distributions for the samples of high and low UV incidence were analysed as it
  was done in Section \ref{sec:completeness}, and the completeness limits found for the two samples
  were qualitatively the same. The magnitude range in which the two samples can be considered
  complete is the same as for the full candidate member sample (Figure \ref{fig:hmag}), hence we do
  not consider that completeness issues in the UV-selected samples may be impacting the results.
Nevertheless, variability surveys sensitive to fainter stars may improve the size and completeness of
the samples towards the fainter stars and help confirming the results presented here. }

Further investigations are needed in order to improve these results, in special in order to complete the 
present analysis for faster rotators. But so far, our results suggest that local incident UV radiation 
may have a role on regulating the AM of YSO. In this sense, evaluating the UV radiation arising from 
massive stars may help to explain differences in the rotational properties of low mass stars in young 
cluster similarly aged presented in the literature.

\section{Conclusions}

In this study we presented the first analysis of rotation properties for low mass stars in the 
young OB association Cygnus OB2. We presented results for stars in the mass range 0.1-1.4$\msun$.
We identified and studied a sample of 1679 stars with signs of periodic variability in their 
NIR light curves, out of a sample of 5083 candidate members. After time-series analysis, we confirm 
periodicity in 1291 stars (25$\%$ of total candidate member sample), but after completeness 
analysis, only 894 of those were considered as reliable period measurements. Since the periodic 
sample is strongly aliased for periods shorter than about 2 days, most of our analysis was
limited to the intermediate (3.14$\leq P\leq$6.28 days) and slow ($P>$6.28 days) rotators, 
with detected fast rotators only in the range 2-3.14 days. The main findings of this work are:

\begin{enumerate}
\item We found periods widely distributed between 0.83 days, and 32.49 days, but due to 
  completeness and contamination issues, we only analysed periods longer than 2 days. The 
  amplitudes of variability for periodic stars were in the range 0.03-1.2 magnitudes; and 
  the masses in the range 0.1-1.4 $M_\odot$.
\item Disk-bearing and non-disked stars are statistically distinct in their rotational properties.
  Even though there is a significant overlap between their period distributions, disk-bearing
  stars rotate on average slower (median period: 7.34 days) than non-disked stars (median period 5.80 
  days). Also, period detection is more common among non-disked stars, than among disked stars.
\item The disk fraction increases as a function of period, except for the very-slow rotator 
  bin ($P>$12.56 days), which has smaller statistics number and seems to suffer from 
  contamination from field stars. The disk fraction varies from $\sim$4$\%$ for fast 
  rotators ($P=$2-3.14 days) 
  to $\sim$13$\%$ for slow rotators ($P=$6.28-12.56 days). This corroborates the results expected
  for disk-locking hypothesis, but when compared with semi-empirical simulations from 
  \citet{2015VasconcelosBouvier} testing disk locking hypothesis, the variation of disk-fraction
  as a function of period is shallower than predicted by the simulations.
\item A period-mass connection was verified, \emph{i.e.}, statistically distinct properties were 
  found for different mass ranges. A median period of 5.7 days was found for the mass range
  $M>0.4M_\odot$,and 6.9 days for $M\leq0.4M_\odot$. However this mass-rotation connection 
  is different from other regions: As is the case of NGC 6530, lower mass stars in CygOB2 rotate
  slower than higher mass stars.
\end{enumerate}

  We also investigated the possibility of a correlation between the incident UV flux arising 
  from O stars in CygOB2 and the rotational properties of low mass stars described in 
  the previous items.

\begin{enumerate}
\setcounter{enumi}{4}
\item We verified that while the distinction in the rotational properties of disk-bearing and non-
  disked stars is stronger in regions with low UV incidence, in regions with high UV incidence it 
  is not possible to distinguish between the two samples. 
\item The increase of the disk fraction with period is stronger for a sample of stars with 
  low local UV incidence, and it is weaker for a sample of stars with high local UV incidence. 
  A maximum disk fraction of 21$\%$, and 7.3$\%$ for slow rotators was found for low and high local 
  UV incidence respectively. 
\item Both low and high local UV incidence samples show a mass-rotation connection with lower mass 
  stars rotating in average slower than higher mass stars. For stars with low local UV incidence,
  a KS-tests gives that this difference is barely significant, but for stars with high local UV 
  incidence, KS-tests shows stronger evidence that the two mass ranges have different rotational 
  period distributions. 
\end{enumerate}

Our results suggest a link between environmental conditions and the rotational evolution of PMS 
stars. However, it is urgent to complement the sample presented in this study for fast rotators 
and lower masses, in order to achieve a better understanding of the rotational scenario in the 
association, and to confirm such suggestions.

\begin{acknowledgements}
This study was part of JR PhD thesis which was granted by CNPq (Conselho Nacional de Desenvolvimento
Cient\'ifico e Tecnol\'ogico) and CAPES (Coordena\c c\~ao de Aperfei\c coamento de Pessoal de N\'ivel
Superior). S.H.P.A acknowledges financial support from CNPq, CAPES, and FAPEMIG.
We thank Bo Reipurth for kindly providing us with the WFCAM/UKIRT dataset used in the present study.
JB acknowledges the support of ANR grant 2011 Blanc SIMI5-6 020 01 Toupies: Towards understanding the
spin evolution of stars (\url{http://ipag.osug.fr/Anr_Toupies/}). 

We also thank Laura Venuti, Francisco Maia, Alana Souza, Jaqueline Vasconcelos for useful comments
over the work development.  

This study has made use of NASA's Astrophysics Data System Bibliographic Services. This research 
did extensive use of TOPCAT software \citep{TOPCAT}.
\end{acknowledgements}

\bibliographystyle{aa}   
\bibliography{bib}

\begin{thebibliography}{113}
\expandafter\ifx\csname natexlab\endcsname\relax\def\natexlab#1{#1}\fi

\bibitem[{{Adams} {et~al.}(2004){Adams}, {Hollenbach}, {Laughlin}, \&
  {Gorti}}]{2004Adams}
{Adams}, F.~C., {Hollenbach}, D., {Laughlin}, G., \& {Gorti}, U. 2004, \apj,
  611, 360

\bibitem[{{Affer} {et~al.}(2013){Affer}, {Micela}, {Favata}, {Flaccomio}, \&
  {Bouvier}}]{Affer2013}
{Affer}, L., {Micela}, G., {Favata}, F., {Flaccomio}, E., \& {Bouvier}, J.
  2013, \mnras, 430, 1433

\bibitem[{{Aihara} {et~al.}(2011){Aihara}, {Allende Prieto}, {An}, {Anderson},
  {Aubourg}, {Balbinot}, {Beers}, {Berlind}, {Bickerton}, {Bizyaev}, {Blanton},
  {Bochanski}, {Bolton}, {Bovy}, {Brandt}, {Brinkmann}, {Brown}, {Brownstein},
  {Busca}, {Campbell}, {Carr}, {Chen}, {Chiappini}, {Comparat}, {Connolly},
  {Cortes}, {Croft}, {Cuesta}, {da Costa}, {Davenport}, {Dawson}, {Dhital},
  {Ealet}, {Ebelke}, {Edmondson}, {Eisenstein}, {Escoffier}, {Esposito},
  {Evans}, {Fan}, {Femen{\'{\i}}a Castell{\'a}}, {Font-Ribera}, {Frinchaboy},
  {Ge}, {Gillespie}, {Gilmore}, {Gonz{\'a}lez Hern{\'a}ndez}, {Gott}, {Gould},
  {Grebel}, {Gunn}, {Hamilton}, {Harding}, {Harris}, {Hawley}, {Hearty}, {Ho},
  {Hogg}, {Holtzman}, {Honscheid}, {Inada}, {Ivans}, {Jiang}, {Johnson},
  {Jordan}, {Jordan}, {Kazin}, {Kirkby}, {Klaene}, {Knapp}, {Kneib},
  {Kochanek}, {Koesterke}, {Kollmeier}, {Kron}, {Lampeitl}, {Lang}, {Le Goff},
  {Lee}, {Lin}, {Long}, {Loomis}, {Lucatello}, {Lundgren}, {Lupton}, {Ma},
  {MacDonald}, {Mahadevan}, {Maia}, {Makler}, {Malanushenko}, {Malanushenko},
  {Mandelbaum}, {Maraston}, {Margala}, {Masters}, {McBride}, {McGehee},
  {McGreer}, {M{\'e}nard}, {Miralda-Escud{\'e}}, {Morrison}, {Mullally},
  {Muna}, {Munn}, {Murayama}, {Myers}, {Naugle}, {Neto}, {Nguyen}, {Nichol},
  {O'Connell}, {Ogando}, {Olmstead}, {Oravetz}, {Padmanabhan},
  {Palanque-Delabrouille}, {Pan}, {Pandey}, {P{\^a}ris}, {Percival},
  {Petitjean}, {Pfaffenberger}, {Pforr}, {Phleps}, {Pichon}, {Pieri}, {Prada},
  {Price-Whelan}, {Raddick}, {Ramos}, {Reyl{\'e}}, {Rich}, {Richards}, {Rix},
  {Robin}, {Rocha-Pinto}, {Rockosi}, {Roe}, {Rollinde}, {Ross}, {Ross},
  {Rossetto}, {S{\'a}nchez}, {Sayres}, {Schlegel}, {Schlesinger}, {Schmidt},
  {Schneider}, {Sheldon}, {Shu}, {Simmerer}, {Simmons}, {Sivarani}, {Snedden},
  {Sobeck}, {Steinmetz}, {Strauss}, {Szalay}, {Tanaka}, {Thakar}, {Thomas},
  {Tinker}, {Tofflemire}, {Tojeiro}, {Tremonti}, {Vandenberg}, {Vargas
  Maga{\~n}a}, {Verde}, {Vogt}, {Wake}, {Wang}, {Weaver}, {Weinberg}, {White},
  {White}, {Yanny}, {Yasuda}, {Yeche}, \& {Zehavi}}]{SDSSDR8}
{Aihara}, H., {Allende Prieto}, C., {An}, D., {et~al.} 2011, \apjs, 193, 29

\bibitem[{{Allard} {et~al.}(2011){Allard}, {Homeier}, \&
  {Freytag}}]{2011Allard}
{Allard}, F., {Homeier}, D., \& {Freytag}, B. 2011, in Astronomical Society of
  the Pacific Conference Series, Vol. 448, 16th Cambridge Workshop on Cool
  Stars, Stellar Systems, and the Sun, ed. C.~{Johns-Krull}, M.~K. {Browning},
  \& A.~A. {West}, 91

\bibitem[{{Allen} {et~al.}(2012){Allen}, {Gutermuth}, {Kryukova}, {Megeath},
  {Pipher}, {Naylor}, {Jeffries}, {Wolk}, {Spitzbart}, \&
  {Muzerolle}}]{2012Allen}
{Allen}, T.~S., {Gutermuth}, R.~A., {Kryukova}, E., {et~al.} 2012, \apj, 750,
  125

\bibitem[{{Baraffe} {et~al.}(1998){Baraffe}, {Chabrier}, {Allard}, \&
  {Hauschildt}}]{1998Baraffe}
{Baraffe}, I., {Chabrier}, G., {Allard}, F., \& {Hauschildt}, P.~H. 1998, \aap,
  337, 403

\bibitem[{{Baraffe} {et~al.}(2015){Baraffe}, {Homeier}, {Allard}, \&
  {Chabrier}}]{2015Baraffe}
{Baraffe}, I., {Homeier}, D., {Allard}, F., \& {Chabrier}, G. 2015, \aap, 577,
  A42

\bibitem[{{Barentsen} {et~al.}(2014){Barentsen}, {Farnhill}, {Drew},
  {Gonz{\'a}lez-Solares}, {Greimel}, {Irwin}, {Miszalski}, {Ruhland}, {Groot},
  {Mampaso}, {Sale}, {Henden}, {Aungwerojwit}, {Barlow}, {Carter}, {Corradi},
  {Drake}, {Eisl{\"o}ffel}, {Fabregat}, {G{\"a}nsicke}, {Gentile Fusillo},
  {Greiss}, {Hales}, {Hodgkin}, {Huckvale}, {Irwin}, {King}, {Knigge},
  {Kupfer}, {Lagadec}, {Lennon}, {Lewis}, {Mohr-Smith}, {Morris}, {Naylor},
  {Parker}, {Phillipps}, {Pyrzas}, {Raddi}, {Roelofs}, {Rodr{\'{\i}}guez-Gil},
  {Sabin}, {Scaringi}, {Steeghs}, {Suso}, {Tata}, {Unruh}, {van Roestel},
  {Viironen}, {Vink}, {Walton}, {Wright}, \& {Zijlstra}}]{IPHAS2014}
{Barentsen}, G., {Farnhill}, H.~J., {Drew}, J.~E., {et~al.} 2014, \mnras, 444,
  3230

\bibitem[{{Beerer} {et~al.}(2010){Beerer}, {Koenig}, {Hora}, {Gutermuth},
  {Bontemps}, {Megeath}, {Schneider}, {Motte}, {Carey}, {Simon}, {Keto},
  {Smith}, {Allen}, {Fazio}, {Kraemer}, {Price}, {Mizuno}, {Adams},
  {Hern{\'a}ndez}, \& {Lucas}}]{SpitzerCygX}
{Beerer}, I.~M., {Koenig}, X.~P., {Hora}, J.~L., {et~al.} 2010, \apj, 720, 679

\bibitem[{{Bell} {et~al.}(2013){Bell}, {Naylor}, {Mayne}, {Jeffries}, \&
  {Littlefair}}]{2013Bell}
{Bell}, C.~P.~M., {Naylor}, T., {Mayne}, N.~J., {Jeffries}, R.~D., \&
  {Littlefair}, S.~P. 2013, \mnras, 434, 806

\bibitem[{{Bell} {et~al.}(2014){Bell}, {Rees}, {Naylor}, {Mayne}, {Jeffries},
  {Mamajek}, \& {Rowe}}]{2014Bell}
{Bell}, C.~P.~M., {Rees}, J.~M., {Naylor}, T., {et~al.} 2014, \mnras, 445, 3496

\bibitem[{{Blaauw}(1964)}]{1964Blaauw}
{Blaauw}, A. 1964, \araa, 2, 213

\bibitem[{{Blaauw} {et~al.}(1959){Blaauw}, {Hiltner}, \&
  {Johnson}}]{1959Blaauw}
{Blaauw}, A., {Hiltner}, W.~A., \& {Johnson}, H.~L. 1959, \apj, 130, 69

\bibitem[{{Bodenheimer}(1995)}]{Bodenheimer}
{Bodenheimer}, P. 1995, \araa, 33, 199

\bibitem[{{Bouvier} {et~al.}(1986){Bouvier}, {Bertout}, {Benz}, \&
  {Mayor}}]{1986jbouvier}
{Bouvier}, J., {Bertout}, C., {Benz}, W., \& {Mayor}, M. 1986, \aap, 165, 110

\bibitem[{{Bouvier} {et~al.}(2003){Bouvier}, {Grankin}, {Alencar}, {Dougados},
  {Fern{\'a}ndez}, {Basri}, {Batalha}, {Guenther}, {Ibrahimov}, {Magakian},
  {Melnikov}, {Petrov}, {Rud}, \& {Zapatero Osorio}}]{2003Bouvier}
{Bouvier}, J., {Grankin}, K.~N., {Alencar}, S.~H.~P., {et~al.} 2003, \aap, 409,
  169

\bibitem[{{Bouvier} {et~al.}(2014){Bouvier}, {Matt}, {Mohanty}, {Scholz},
  {Stassun}, \& {Zanni}}]{BouvierMatt2013ReviewII}
{Bouvier}, J., {Matt}, S.~P., {Mohanty}, S., {et~al.} 2014, Protostars and
  Planets VI, 433

\bibitem[{{Bouy} {et~al.}(2013){Bouy}, {Bertin}, {Moraux}, {Cuillandre},
  {Bouvier}, {Barrado}, {Solano}, \& {Bayo}}]{DANCe1st}
{Bouy}, H., {Bertin}, E., {Moraux}, E., {et~al.} 2013, \aap, 554, A101

\bibitem[{{Cardelli} {et~al.}(1989){Cardelli}, {Clayton}, \&
  {Mathis}}]{1989Cardelli+}
{Cardelli}, J.~A., {Clayton}, G.~C., \& {Mathis}, J.~S. 1989, \apj, 345, 245

\bibitem[{{Carpenter} {et~al.}(2001){Carpenter}, {Hillenbrand}, \&
  {Skrutskie}}]{2001Carpenter}
{Carpenter}, J.~M., {Hillenbrand}, L.~A., \& {Skrutskie}, M.~F. 2001, \aj, 121,
  3160

\bibitem[{{Casali} {et~al.}(2007){Casali}, {Adamson}, {Alves de Oliveira},
  {Almaini}, {Burch}, {Chuter}, {Elliot}, {Folger}, {Foucaud}, {Hambly},
  {Hastie}, {Henry}, {Hirst}, {Irwin}, {Ives}, {Lawrence}, {Laidlaw}, {Lee},
  {Lewis}, {Lunney}, {McLay}, {Montgomery}, {Pickup}, {Read}, {Rees}, {Robson},
  {Sekiguchi}, {Vick}, {Warren}, \& {Woodward}}]{wfcam}
{Casali}, M., {Adamson}, A., {Alves de Oliveira}, C., {et~al.} 2007, \aap, 467,
  777

\bibitem[{{Cieza} \& {Baliber}(2007)}]{2007CiezaBaliber}
{Cieza}, L. \& {Baliber}, N. 2007, \apj, 671, 605

\bibitem[{{Clarke}(2002)}]{Clarke2002}
{Clarke}, D. 2002, \aap, 386, 763

\bibitem[{{Cody} \& {Hillenbrand}(2010)}]{2010Cody}
{Cody}, A.~M. \& {Hillenbrand}, L.~A. 2010, \apjs, 191, 389

\bibitem[{{Cody} {et~al.}(2014){Cody}, {Stauffer}, {Baglin}, {Micela},
  {Rebull}, {Flaccomio}, {Morales-Calder{\'o}n}, {Aigrain}, {Bouvier},
  {Hillenbrand}, {Gutermuth}, {Song}, {Turner}, {Alencar}, {Zwintz},
  {Plavchan}, {Carpenter}, {Findeisen}, {Carey}, {Terebey}, {Hartmann},
  {Calvet}, {Teixeira}, {Vrba}, {Wolk}, {Covey}, {Poppenhaeger}, {G{\"u}nther},
  {Forbrich}, {Whitney}, {Affer}, {Herbst}, {Hora}, {Barrado}, {Holtzman},
  {Marchis}, {Wood}, {Medeiros Guimar{\~a}es}, {Lillo Box}, {Gillen},
  {McQuillan}, {Espaillat}, {Allen}, {D'Alessio}, \& {Favata}}]{2014Cody}
{Cody}, A.~M., {Stauffer}, J., {Baglin}, A., {et~al.} 2014, \aj, 147, 82

\bibitem[{{Cohen} {et~al.}(2004){Cohen}, {Herbst}, \& {Williams}}]{2004Cohen}
{Cohen}, R.~E., {Herbst}, W., \& {Williams}, E.~C. 2004, \aj, 127, 1602

\bibitem[{{Comer{\'o}n} {et~al.}(2002){Comer{\'o}n}, {Pasquali}, {Rodighiero},
  {Stanishev}, {De Filippis}, {L{\'o}pez Mart{\'{\i}}}, {G{\'a}lvez Ortiz},
  {Stankov}, \& {Gredel}}]{comeron2002+}
{Comer{\'o}n}, F., {Pasquali}, A., {Rodighiero}, G., {et~al.} 2002, \aap, 389,
  874

\bibitem[{{Covey} {et~al.}(2007){Covey}, {Ivezi{\'c}}, {Schlegel},
  {Finkbeiner}, {Padmanabhan}, {Lupton}, {Ag{\"u}eros}, {Bochanski}, {Hawley},
  {West}, {Seth}, {Kimball}, {Gogarten}, {Claire}, {Haggard}, {Kaib},
  {Schneider}, \& {Sesar}}]{Covey}
{Covey}, K.~R., {Ivezi{\'c}}, {\v Z}., {Schlegel}, D., {et~al.} 2007, \aj, 134,
  2398

\bibitem[{{Cutri} {et~al.}(2003){Cutri}, {Skrutskie}, {van Dyk}, {Beichman},
  {Carpenter}, {Chester}, {Cambresy}, {Evans}, {Fowler}, {Gizis}, {Howard},
  {Huchra}, {Jarrett}, {Kopan}, {Kirkpatrick}, {Light}, {Marsh}, {McCallon},
  {Schneider}, {Stiening}, {Sykes}, {Weinberg}, {Wheaton}, {Wheelock}, \&
  {Zacarias}}]{2MASS}
{Cutri}, R.~M., {Skrutskie}, M.~F., {van Dyk}, S., {et~al.} 2003, {2MASS All
  Sky Catalog of point sources.}

\bibitem[{{Dahm}(2008)}]{2008Dahm}
{Dahm}, S.~E. 2008, {The Young Cluster and Star Forming Region NGC 2264}, ed.
  B.~{Reipurth}, 966

\bibitem[{{Damiani} {et~al.}(2004){Damiani}, {Flaccomio}, {Micela},
  {Sciortino}, {Harnden}, \& {Murray}}]{2004Damiani}
{Damiani}, F., {Flaccomio}, E., {Micela}, G., {et~al.} 2004, \apj, 608, 781

\bibitem[{{Doi} {et~al.}(2010){Doi}, {Tanaka}, {Fukugita}, {Gunn}, {Yasuda},
  {Ivezi{\'c}}, {Brinkmann}, {de Haars}, {Kleinman}, {Krzesinski}, \& {French
  Leger}}]{2010Doi}
{Doi}, M., {Tanaka}, M., {Fukugita}, M., {et~al.} 2010, \aj, 139, 1628

\bibitem[{{Drew} {et~al.}(2005){Drew}, {Greimel}, {Irwin}, {Aungwerojwit},
  {Barlow}, {Corradi}, {Drake}, {G{\"a}nsicke}, {Groot}, {Hales}, {Hopewell},
  {Irwin}, {Knigge}, {Leisy}, {Lennon}, {Mampaso}, {Masheder}, {Matsuura},
  {Morales-Rueda}, {Morris}, {Parker}, {Phillipps}, {Rodriguez-Gil}, {Roelofs},
  {Skillen}, {Sokoloski}, {Steeghs}, {Unruh}, {Viironen}, {Vink}, {Walton},
  {Witham}, {Wright}, {Zijlstra}, \& {Zurita}}]{IPHAS2005}
{Drew}, J.~E., {Greimel}, R., {Irwin}, M.~J., {et~al.} 2005, \mnras, 362, 753

\bibitem[{{Drew} {et~al.}(2008){Drew}, {Greimel}, {Irwin}, \&
  {Sale}}]{Drew2008+}
{Drew}, J.~E., {Greimel}, R., {Irwin}, M.~J., \& {Sale}, S.~E. 2008, \mnras,
  386, 1761

\bibitem[{{Gallet} \& {Bouvier}(2013)}]{2013Gallet}
{Gallet}, F. \& {Bouvier}, J. 2013, \aap, 556, A36

\bibitem[{{Ghosh} \& {Lamb}(1979)}]{GhoshLamb1979}
{Ghosh}, P. \& {Lamb}, F.~K. 1979, \apj, 234, 296

\bibitem[{{Gregory} {et~al.}(2012){Gregory}, {Donati}, {Morin}, {Hussain},
  {Mayne}, {Hillenbrand}, \& {Jardine}}]{2012Gregory}
{Gregory}, S.~G., {Donati}, J.-F., {Morin}, J., {et~al.} 2012, \apj, 755, 97

\bibitem[{{Guarcello} {et~al.}(2016){Guarcello}, {Drake}, {Wright},
  {Albacete-Colombo}, {Clarke}, {Ercolano}, {Flaccomio}, {Kashyap}, {Micela},
  {Naylor}, {Schneider}, {Sciortino}, \& {Vink}}]{gdw16}
{Guarcello}, M.~G., {Drake}, J.~J., {Wright}, N.~J., {et~al.} 2016, ArXiv
  e-prints [\eprint[arXiv]{1605.01773}]

\bibitem[{{Guarcello} {et~al.}(2013){Guarcello}, {Drake}, {Wright}, {Drew},
  {Gutermuth}, {Hora}, {Naylor}, {Aldcroft}, {Fruscione},
  {Garc{\'{\i}}a-Alvarez}, {Kashyap}, \& {King}}]{gdw13}
{Guarcello}, M.~G., {Drake}, J.~J., {Wright}, N.~J., {et~al.} 2013, \apj, 773,
  135

\bibitem[{{Guarcello} {et~al.}(2015){Guarcello}, {Drake}, {Wright}, {Naylor},
  {Flaccomio}, {Kashyap}, \& {Garcia-Alvarez}}]{gdw15}
{Guarcello}, M.~G., {Drake}, J.~J., {Wright}, N.~J., {et~al.} 2015, ArXiv
  e-prints [\eprint[arXiv]{1501.03761}]

\bibitem[{{Guarcello} {et~al.}(2010){Guarcello}, {Micela}, {Peres},
  {Prisinzano}, \& {Sciortino}}]{2010Guarcello}
{Guarcello}, M.~G., {Micela}, G., {Peres}, G., {Prisinzano}, L., \&
  {Sciortino}, S. 2010, \aap, 521, A61

\bibitem[{{Guarcello} {et~al.}(2007){Guarcello}, {Prisinzano}, {Micela},
  {Damiani}, {Peres}, \& {Sciortino}}]{2007Guarcello}
{Guarcello}, M.~G., {Prisinzano}, L., {Micela}, G., {et~al.} 2007, \aap, 462,
  245

\bibitem[{{Guarcello} {et~al.}(2012){Guarcello}, {Wright}, {Drake},
  {Garc{\'{\i}}a-Alvarez}, {Drew}, {Aldcroft}, \& {Kashyap}}]{Guarcello2012+}
{Guarcello}, M.~G., {Wright}, N.~J., {Drake}, J.~J., {et~al.} 2012, \apjs, 202,
  19

\bibitem[{{Habing}(1968)}]{1968Habing}
{Habing}, H.~J. 1968, \bain, 19, 421

\bibitem[{{Hanson}(2003)}]{hanson2003}
{Hanson}, M.~M. 2003, \apj, 597, 957

\bibitem[{{Henderson} {et~al.}(2011){Henderson}, {Stanek}, {Pejcha}, \&
  {Prieto}}]{hend2011+}
{Henderson}, C.~B., {Stanek}, K.~Z., {Pejcha}, O., \& {Prieto}, J.~L. 2011,
  \apjs, 194, 27

\bibitem[{{Henderson} \& {Stassun}(2012)}]{2012HendersonStassun}
{Henderson}, C.~B. \& {Stassun}, K.~G. 2012, \apj, 747, 51

\bibitem[{{Herbst} {et~al.}(2001){Herbst}, {Bailer-Jones}, \&
  {Mundt}}]{2001Herbst}
{Herbst}, W., {Bailer-Jones}, C.~A.~L., \& {Mundt}, R. 2001, \apjl, 554, L197

\bibitem[{{Herbst} {et~al.}(2002){Herbst}, {Bailer-Jones}, {Mundt},
  {Meisenheimer}, \& {Wackermann}}]{2002Herbst}
{Herbst}, W., {Bailer-Jones}, C.~A.~L., {Mundt}, R., {Meisenheimer}, K., \&
  {Wackermann}, R. 2002, \aap, 396, 513

\bibitem[{{Herbst} \& {Mundt}(2005)}]{2005Herbst}
{Herbst}, W. \& {Mundt}, R. 2005, \apj, 633, 967

\bibitem[{{Hewett} {et~al.}(2006){Hewett}, {Warren}, {Leggett}, \&
  {Hodgkin}}]{filtroukirt}
{Hewett}, P.~C., {Warren}, S.~J., {Leggett}, S.~K. T.~d., \& {Hodgkin}, S.~T.
  2006, \mnras, 367, 454

\bibitem[{{Hillenbrand} {et~al.}(1998){Hillenbrand}, {Strom}, {Calvet},
  {Merrill}, {Gatley}, {Makidon}, {Meyer}, \& {Skrutskie}}]{1998Hillenbrand}
{Hillenbrand}, L.~A., {Strom}, S.~E., {Calvet}, N., {et~al.} 1998, \aj, 116,
  1816

\bibitem[{{Hodgkin} {et~al.}(2009){Hodgkin}, {Irwin}, {Hewett}, \&
  {Warren}}]{wfcamcalibration}
{Hodgkin}, S.~T., {Irwin}, M.~J., {Hewett}, P.~C., \& {Warren}, S.~J. 2009,
  \mnras, 394, 675

\bibitem[{{Horne} \& {Baliunas}(1986)}]{horne}
{Horne}, J.~H. \& {Baliunas}, S.~L. 1986, \apj, 302, 757

\bibitem[{{Irwin} \& {Bouvier}(2009)}]{IrwinBouvier2009}
{Irwin}, J. \& {Bouvier}, J. 2009, in IAU Symposium, Vol. 258, IAU Symposium,
  ed. E.~E. {Mamajek}, D.~R. {Soderblom}, \& R.~F.~G. {Wyse}, 363--374

\bibitem[{{Irwin} {et~al.}(2008){Irwin}, {Hodgkin}, {Aigrain}, {Bouvier},
  {Hebb}, {Irwin}, \& {Moraux}}]{Irwin2008+}
{Irwin}, J., {Hodgkin}, S., {Aigrain}, S., {et~al.} 2008, \mnras, 384, 675

\bibitem[{{Johnstone} {et~al.}(1998){Johnstone}, {Hollenbach}, \&
  {Bally}}]{1998Johnstone}
{Johnstone}, D., {Hollenbach}, D., \& {Bally}, J. 1998, \apj, 499, 758

\bibitem[{{Joy}(1945)}]{1945Joy}
{Joy}, A.~H. 1945, \apj, 102, 168

\bibitem[{{Kashyap}(2017)}]{2016Kashyap}
{Kashyap}, V.~L. 2017

\bibitem[{{Kenyon} \& {Hartmann}(1995)}]{1995KenyonHartmann}
{Kenyon}, S.~J. \& {Hartmann}, L. 1995, \apjs, 101, 117

\bibitem[{{Kiminki} {et~al.}(2015){Kiminki}, {Kobulnicky}, {Vargas
  {\'A}lvarez}, {Alexander}, \& {Lundquist}}]{2015Kiminki}
{Kiminki}, D.~C., {Kobulnicky}, H.~A., {Vargas {\'A}lvarez}, C.~A.,
  {Alexander}, M.~J., \& {Lundquist}, M.~J. 2015, \apj, 811, 85

\bibitem[{{Kn{\"o}dlseder}(2000)}]{knod2000}
{Kn{\"o}dlseder}, J. 2000, \aap, 360, 539

\bibitem[{{Koenigl}(1991)}]{Koenigl1991}
{Koenigl}, A. 1991, \apjl, 370, L39

\bibitem[{{Lafler} \& {Kinman}(1965)}]{65LK}
{Lafler}, J. \& {Kinman}, T.~D. 1965, \apjs, 11, 216

\bibitem[{{Lamm} {et~al.}(2005){Lamm}, {Mundt}, {Bailer-Jones}, \&
  {Herbst}}]{2005Lamm}
{Lamm}, M.~H., {Mundt}, R., {Bailer-Jones}, C.~A.~L., \& {Herbst}, W. 2005,
  \aap, 430, 1005

\bibitem[{{Littlefair} {et~al.}(2005){Littlefair}, {Naylor}, {Burningham}, \&
  {Jeffries}}]{Littlefair2005}
{Littlefair}, S.~P., {Naylor}, T., {Burningham}, B., \& {Jeffries}, R.~D. 2005,
  \mnras, 358, 341

\bibitem[{{Littlefair} {et~al.}(2010){Littlefair}, {Naylor}, {Mayne},
  {Saunders}, \& {Jeffries}}]{Littlefair2010+}
{Littlefair}, S.~P., {Naylor}, T., {Mayne}, N.~J., {Saunders}, E.~S., \&
  {Jeffries}, R.~D. 2010, \mnras, 403, 545

\bibitem[{{Lomb}(1976)}]{lomb}
{Lomb}, N.~R. 1976, \apss, 39, 447

\bibitem[{{Lucas} {et~al.}(2008){Lucas}, {Hoare}, {Longmore}, {Schr{\"o}der},
  {Davis}, {Adamson}, {Bandyopadhyay}, {de Grijs}, {Smith}, {Gosling},
  {Mitchison}, {G{\'a}sp{\'a}r}, {Coe}, {Tamura}, {Parker}, {Irwin}, {Hambly},
  {Bryant}, {Collins}, {Cross}, {Evans}, {Gonzalez-Solares}, {Hodgkin},
  {Lewis}, {Read}, {Riello}, {Sutorius}, {Lawrence}, {Drew}, {Dye}, \&
  {Thompson}}]{UKIDSSGPS2008}
{Lucas}, P.~W., {Hoare}, M.~G., {Longmore}, A., {et~al.} 2008, \mnras, 391, 136

\bibitem[{{Massey} \& {Thompson}(1991)}]{MassThomp91}
{Massey}, P. \& {Thompson}, A.~B. 1991, \aj, 101, 1408

\bibitem[{{Matt} \& {Pudritz}(2005)}]{2005Matta}
{Matt}, S. \& {Pudritz}, R.~E. 2005, \apjl, 632, L135

\bibitem[{{Matt} {et~al.}(2012){Matt}, {Pinz{\'o}n}, {Greene}, \&
  {Pudritz}}]{Matt2012}
{Matt}, S.~P., {Pinz{\'o}n}, G., {Greene}, T.~P., \& {Pudritz}, R.~E. 2012,
  \apj, 745, 101

\bibitem[{{Mayne} \& {Naylor}(2008)}]{2008MayneNaylor}
{Mayne}, N.~J. \& {Naylor}, T. 2008, \mnras, 386, 261

\bibitem[{{McGinnis} {et~al.}(2015){McGinnis}, {Alencar}, {Guimar{\~a}es},
  {Sousa}, {Stauffer}, {Bouvier}, {Rebull}, {Fonseca}, {Venuti}, {Hillenbrand},
  {Cody}, {Teixeira}, {Aigrain}, {Favata}, {F{\H u}r{\'e}sz}, {Vrba},
  {Flaccomio}, {Turner}, {Gameiro}, {Dougados}, {Herbst},
  {Morales-Calder{\'o}n}, \& {Micela}}]{2015McGinnis}
{McGinnis}, P.~T., {Alencar}, S.~H.~P., {Guimar{\~a}es}, M.~M., {et~al.} 2015,
  \aap, 577, A11

\bibitem[{{Moraux} {et~al.}(2013){Moraux}, {Artemenko}, {Bouvier}, {Irwin},
  {Ibrahimov}, {Magakian}, {Grankin}, {Nikogossian}, {Cardoso}, {Hodgkin},
  {Aigrain}, \& {Movsessian}}]{2013Moraux}
{Moraux}, E., {Artemenko}, S., {Bouvier}, J., {et~al.} 2013, \aap, 560, A13

\bibitem[{{Naylor}(2009)}]{2009Naylor}
{Naylor}, T. 2009, \mnras, 399, 432

\bibitem[{{Pasquali} {et~al.}(2002){Pasquali}, {Comer{\'o}n}, {Gredel},
  {Torra}, \& {Figueras}}]{Pasquali2002+}
{Pasquali}, A., {Comer{\'o}n}, F., {Gredel}, R., {Torra}, J., \& {Figueras}, F.
  2002, \aap, 396, 533

\bibitem[{{Plavchan} {et~al.}(2008){Plavchan}, {Jura}, {Kirkpatrick}, {Cutri},
  \& {Gallagher}}]{2008Plavchan}
{Plavchan}, P., {Jura}, M., {Kirkpatrick}, J.~D., {Cutri}, R.~M., \&
  {Gallagher}, S.~C. 2008, \apjs, 175, 191

\bibitem[{Press {et~al.}(1992)Press, Flannery, Teukolsky, \&
  Veterling}]{numericalr}
Press, W.~H., Flannery, B.~P., Teukolsky, S.~A., \& Veterling, W.~T. 1992,
  Numerical Recipes in C: the art of scientific computing, 2nd edn. (Cambridge:
  Cambridge University Press)

\bibitem[{{Rauw} {et~al.}(2015){Rauw}, {Naz{\'e}}, {Wright}, {Drake},
  {Guarcello}, {Prinja}, {Peck}, {Albacete Colombo}, {Herrero}, {Kobulnicky},
  {Sciortino}, \& {Vink}}]{2015Rauw}
{Rauw}, G., {Naz{\'e}}, Y., {Wright}, N.~J., {et~al.} 2015, \apjs, 221, 1

\bibitem[{{Rebull} {et~al.}(2004){Rebull}, {Wolff}, \& {Strom}}]{2004Rebull}
{Rebull}, L.~M., {Wolff}, S.~C., \& {Strom}, S.~E. 2004, \aj, 127, 1029

\bibitem[{{Reddish} {et~al.}(1966){Reddish}, {Lawrence}, \&
  {Pratt}}]{Reddish66+}
{Reddish}, V., {Lawrence}, L.~C., \& {Pratt}, N.~M. 1966, Publications of the
  Royal Observatory of Edinburgh, 5, 111

\bibitem[{{Rice} {et~al.}(2015){Rice}, {Reipurth}, {Wolk}, {Vaz}, \&
  {Cross}}]{2015Rice}
{Rice}, T.~S., {Reipurth}, B., {Wolk}, S.~J., {Vaz}, L.~P., \& {Cross},
  N.~J.~G. 2015, \aj, 150, 132

\bibitem[{{Rodr{\'{\i}}guez-Ledesma} {et~al.}(2009){Rodr{\'{\i}}guez-Ledesma},
  {Mundt}, \& {Eisl{\"o}ffel}}]{RodriguezLedesma2009}
{Rodr{\'{\i}}guez-Ledesma}, M.~V., {Mundt}, R., \& {Eisl{\"o}ffel}, J. 2009,
  \aap, 502, 883

\bibitem[{{Romanova} {et~al.}(2009){Romanova}, {Ustyugova}, {Koldoba}, \&
  {Lovelace}}]{2009Romanova}
{Romanova}, M.~M., {Ustyugova}, G.~V., {Koldoba}, A.~V., \& {Lovelace},
  R.~V.~E. 2009, \mnras, 399, 1802

\bibitem[{{Roquette}(2017)}]{2017Roquette}
{Roquette}, J. 2017, in preparation

\bibitem[{{Saunders} {et~al.}(2006){Saunders}, {Naylor}, \&
  {Allan}}]{2006Saunders}
{Saunders}, E.~S., {Naylor}, T., \& {Allan}, A. 2006, Astronomische
  Nachrichten, 327, 783

\bibitem[{{Scargle}(1982)}]{scargle2}
{Scargle}, J.~D. 1982, \apj, 263, 835

\bibitem[{{Schlegel} {et~al.}(1998){Schlegel}, {Finkbeiner}, \&
  {Davis}}]{1998Schlegel}
{Schlegel}, D.~J., {Finkbeiner}, D.~P., \& {Davis}, M. 1998, \apj, 500, 525

\bibitem[{{Scholz} \& {Eisl{\"o}ffel}(2004)}]{2004Scholz}
{Scholz}, A. \& {Eisl{\"o}ffel}, J. 2004, \aap, 419, 249

\bibitem[{{Scholz} {et~al.}(2009){Scholz}, {Xu}, {Jayawardhana}, {Wood},
  {Eisl{\"o}ffel}, \& {Quinn}}]{2009Scholz}
{Scholz}, A., {Xu}, X., {Jayawardhana}, R., {et~al.} 2009, \mnras, 398, 873

\bibitem[{{Siess} {et~al.}(2000){Siess}, {Dufour}, \& {Forestini}}]{2000Siess}
{Siess}, L., {Dufour}, E., \& {Forestini}, M. 2000, \aap, 358, 593

\bibitem[{{Sousa} {et~al.}(2016){Sousa}, {Alencar}, {Bouvier}, {Stauffer},
  {Venuti}, {Hillenbrand}, {Cody}, {Teixeira}, {Guimar{\~a}es}, {McGinnis},
  {Rebull}, {Flaccomio}, {F{\"u}r{\'e}sz}, {Micela}, \& {Gameiro}}]{2016Sousa}
{Sousa}, A.~P., {Alencar}, S.~H.~P., {Bouvier}, J., {et~al.} 2016, \aap, 586,
  A47

\bibitem[{{Stauffer} {et~al.}(2014){Stauffer}, {Cody}, {Baglin}, {Alencar},
  {Rebull}, {Hillenbrand}, {Venuti}, {Turner}, {Carpenter}, {Plavchan},
  {Findeisen}, {Carey}, {Terebey}, {Morales-Calder{\'o}n}, {Bouvier}, {Micela},
  {Flaccomio}, {Song}, {Gutermuth}, {Hartmann}, {Calvet}, {Whitney}, {Barrado},
  {Vrba}, {Covey}, {Herbst}, {Furesz}, {Aigrain}, \& {Favata}}]{2014Stauffer}
{Stauffer}, J., {Cody}, A.~M., {Baglin}, A., {et~al.} 2014, \aj, 147, 83

\bibitem[{{Stetson}(1996)}]{Stet1996}
{Stetson}, P.~B. 1996, \pasp, 108, 851

\bibitem[{{Sung} {et~al.}(2000){Sung}, {Chun}, \& {Bessell}}]{2000Sung}
{Sung}, H., {Chun}, M.-Y., \& {Bessell}, M.~S. 2000, \aj, 120, 333

\bibitem[{{Tanner}(1948)}]{48Tanner}
{Tanner}, R.~W. 1948, \jrasc, 42, 177

\bibitem[{{Taylor}(2005)}]{TOPCAT}
{Taylor}, M.~B. 2005, in Astronomical Society of the Pacific Conference Series,
  Vol. 347, Astronomical Data Analysis Software and Systems XIV, ed.
  P.~{Shopbell}, M.~{Britton}, \& R.~{Ebert}, 29

\bibitem[{{Tognelli} {et~al.}(2012){Tognelli}, {Degl'Innocenti}, \& {Prada
  Moroni}}]{2012Tognelli}
{Tognelli}, E., {Degl'Innocenti}, S., \& {Prada Moroni}, P.~G. 2012, Memorie
  della Societa Astronomica Italiana Supplementi, 22, 225

\bibitem[{{Tognelli} {et~al.}(2011){Tognelli}, {Prada Moroni}, \&
  {Degl'Innocenti}}]{2011Tognelli}
{Tognelli}, E., {Prada Moroni}, P.~G., \& {Degl'Innocenti}, S. 2011, \aap, 533,
  A109

\bibitem[{{Torres-Dodgen} {et~al.}(1991){Torres-Dodgen}, {Carroll}, \&
  {Tapia}}]{torres91+}
{Torres-Dodgen}, A., {Carroll}, M., \& {Tapia}, M. 1991, \mnras, 249, 1

\bibitem[{{Vasconcelos} \& {Bouvier}(2015)}]{2015VasconcelosBouvier}
{Vasconcelos}, M.~J. \& {Bouvier}, J. 2015, \aap, 578, A89

\bibitem[{{Venuti} {et~al.}(2016){Venuti}, {Bouvier}, {Cody}, {Stauffer},
  {Micela}, {Rebull}, {Alencar}, {Sousa}, {Hillenbrand}, \&
  {Flaccomio}}]{2016Venuti}
{Venuti}, L., {Bouvier}, J., {Cody}, A.~M., {et~al.} 2016, ArXiv e-prints
  [\eprint[arXiv]{1610.08811}]

\bibitem[{{Vink} {et~al.}(2008){Vink}, {Drew}, {Steeghs}, {Wright}, {Martin},
  {G{\"a}nsicke}, {Greimel}, \& {Drake}}]{2008Vink}
{Vink}, J.~S., {Drew}, J.~E., {Steeghs}, D., {et~al.} 2008, \mnras, 387, 308

\bibitem[{{Vogel} \& {Kuhi}(1981)}]{1982Vogel}
{Vogel}, S.~N. \& {Kuhi}, L.~V. 1981, \apj, 245, 960

\bibitem[{{Walborn} {et~al.}(2002){Walborn}, {Howarth}, {Lennon}, {Massey},
  {Oey}, {Moffat}, {Skalkowski}, {Morrell}, {Drissen}, \&
  {Parker}}]{Walborn2002}
{Walborn}, N.~R., {Howarth}, I.~D., {Lennon}, D.~J., {et~al.} 2002, \aj, 123,
  2754

\bibitem[{{Wilking} {et~al.}(2001){Wilking}, {Bontemps}, {Schuler}, {Greene},
  \& {Andr{\'e}}}]{Wilking2001+}
{Wilking}, B.~A., {Bontemps}, S., {Schuler}, R.~E., {Greene}, T.~P., \&
  {Andr{\'e}}, P. 2001, \apj, 551, 357

\bibitem[{{Wright} {et~al.}(2016){Wright}, {Bouy}, {Drew}, {Sarro}, {Bertin},
  {Cuillandre}, \& {Barrado}}]{2016WrightDance}
{Wright}, N.~J., {Bouy}, H., {Drew}, J.~E., {et~al.} 2016, \mnras, 460, 2593

\bibitem[{{Wright} \& {Drake}(2009)}]{wrightdrake2009}
{Wright}, N.~J. \& {Drake}, J.~J. 2009, \apjs, 184, 84

\bibitem[{{Wright} {et~al.}(2010){Wright}, {Drake}, {Drew}, \&
  {Vink}}]{wright2010+}
{Wright}, N.~J., {Drake}, J.~J., {Drew}, J.~E., \& {Vink}, J.~S. 2010, \apj,
  713, 871

\bibitem[{{Wright} {et~al.}(2014){Wright}, {Drake}, {Guarcello}, {Aldcroft},
  {Kashyap}, {Damiani}, {DePasquale}, \& {Fruscione}}]{Wright2014b+}
{Wright}, N.~J., {Drake}, J.~J., {Guarcello}, M.~G., {et~al.} 2014, ArXiv
  e-prints [\eprint[arXiv]{1408.6579}]

\bibitem[{{Wright} {et~al.}(2015){Wright}, {Drew}, \&
  {Mohr-Smith}}]{2015WrightMassive}
{Wright}, N.~J., {Drew}, J.~E., \& {Mohr-Smith}, M. 2015, \mnras, 449, 741

\bibitem[{{Zanni} \& {Ferreira}(2013)}]{2013Zanni}
{Zanni}, C. \& {Ferreira}, J. 2013, \aap, 550, A99

\end{thebibliography}

\begin{appendix}
\
\section{\textbf{Other Possible Sources of Contamination To The Periodic Sample}}
\label{app:contamination}

\subsection{\textbf{Contamination by Periods Outside the Observations Resolution}}
\label{sec:contamination}
\textbf{
Given the evidence of high contamination in the shorter periods bin (Figures \ref{fig:completeness}
and \ref{fig:completeness1}), we performed some extra simulations focusing on periods outside the 
period range considered.
}
\paragraph{\textbf{Longer Periods:}} \textbf{
First, we focused on the possible contamination due to periods longer
than 20 days.  We repeated the simulation described in Section \ref{sec:simulations}, but this time
considering periods ($P_{\mathrm{in}}$) between 20 and 100 days with amplitudes between 0.015 and 0.5
mag. For those specific simulations, we kept the filtering for FAP, RL- and S-statistics, but we did
not perform any further cleaning for periods around 1day. The recovery success on this range is only
$\sim$12$\%$ in a sample of 78000 synthetic periodic light curves for the amplitude range considered.
For the 680 (9$\%$) synthetic light curves in contaminant sample (\emph{i.e.}, synthetic light curves selected as periodic, but with
$P_{\mathrm{out}}\neq P_{\mathrm{in}}$), we analysed the distribution of measured periods
($P_{\mathrm{out}}$), and verified that $\sim$98$\%$ of the contaminant sample had $P_{\mathrm{out}}$
around 1$\pm$0.2 days. This justify the filtering for periods in the range 0.92-1.08 days applied in
Section \ref{sec:periodic}.
}

\begin{figure}[tb]
   \centering
  \includegraphics[width=0.75\columnwidth]{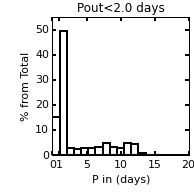}
  \caption{\label{fig:shortperiod}Period distribution histogram for the input periods for synthetic 
  light curves measured  P$_{\mathrm{out}}<$2.0 days (2 shorter period bins in Figure \ref{fig:completeness1}) 
 after the automatized period search procedure.} 
\end{figure}

 \paragraph{\textbf{
 Shorter Periods:}} \textbf{
Second, we focus on the contamination due to periods shorter than our 
 data-sampling is capable of resolving. For the set of simulations presented in Figure 
 \ref{fig:completeness1}, we analysed the input characteristics of the synthetic light curves.
}
\textbf{
 Figure \ref{fig:shortperiod} shows the input period ($P_{\mathrm{in}}$) distribution for the synthetic
 light curves with  P$_{\mathrm{out}}<2$ days, which compose the two shortest period bins in Figure
 \ref{fig:completeness1}. The y-axis shows the fraction of contaminants with $P_\mathrm{out}<2$ days.
 From this plot, we can 
 estimate that $\sim$15$\%$ of the contamination arises from aliased periods smaller than 1 day, and 
 $\sim$49$\%$ of the contamination arises from unresolved periods between 1 and 2 days. Despite the fact 
 that these short period aliases account for $\sim$64$\%$ of the contamination, there is a lower level 
 contamination arising from all period bins larger than 2 days up to $\sim$15 days, which adds up 
 $\sim$36$\%$ of contamination. Since it is not possible to untangle the two contaminant sources, we 
 conclude that the periodic sample with $P<2.0$ days suffers from strong contamination, and it should not be 
 used in the further analysis. 
}
\subsection{Contamination Due to Non-Periodic Variability}

\textbf{We performed additional Monte Carlo simulation for non-periodic waveforms and applied the automatized
period search to the non-periodic sample. For each synthetic light curve, a random amplitude was
generated and two possible waveforms were randomly assigned. The first waveform assumes that the
light-curve is a straight line with slope defined as the ratio between the randomly selected
amplitude and the total number of observations. The second wave form was composed of random numbers
generated between 0.015 and the amplitude value.} 

\textbf{A total of 78000 synthetic light curves were generated this way, with amplitudes between
  0.015 and 1.5 mag.   
While $\sim49\%$ of the sample presented periodogram power peaks in the three
bands with a power higher enough to be automatically selected as periodic, they were however not
wrongly selected as periodic, since most of them are distributed around 1 day, and were ruled out from the periodic
selection. The few objects with periods outside this range were ruled out by the RL-statistics or Saunder-Statistics
filters.
  Finally, after applying the automatized period search procedure to this sample, we
  concluded that the procedure is robust enough to avoid any contamination due to these non-periodic
waveforms in our periodic sampe. }
\end{appendix}


\end{document}